\documentclass[10pt,twoside,twocolumn,british,english,preprintnumbers,superscriptaddress,nofootinbib]{revtex4-1}
\usepackage{amsmath}
\usepackage{amssymb}
\usepackage{graphicx}
\usepackage{babel}
\usepackage{hyperref}
\def\be{\begin{equation}}
\def\ee{\end{equation}}
\def\bea{\begin{eqnarray}}
\def\eea{\end{eqnarray}}
\def\d{\mathrm d}
\def\ov{\overline}

\begin{document}

\title{Evading the Lyth Bound in Hybrid Natural Inflation}

\author{A. Hebecker}
\email{A.Hebecker@ThPhys.Uni-Heidelberg.de}
\author{S. C. Kraus}
\email{S.Kraus@ThPhys.Uni-Heidelberg.de}
\affiliation{Institute for Theoretical Physics, Heidelberg University, D-69120, Germany}

\author{A. Westphal}
\email{Alexander.Westphal@desy.de}
\affiliation{Deutsches Elektronen-Synchrotron DESY, Theory Group, D-22603 Hamburg, Germany}
\affiliation{Kavli Institute for Theoretical Physics, Santa Barbara, California 93106, USA}

\begin{abstract}
Generically, the gravitational-wave or tensor-mode contribution to the 
primordial curvature spectrum of inflation is tiny if the field-range 
of the inflaton is much smaller than the Planck scale. We show that
this pessimistic conclusion is naturally avoided in a rather broad class 
of small-field models. More specifically, we consider models where 
an axion-like shift symmetry keeps the inflaton potential flat (up to
non-perturbative cosine-shaped modulations), but inflation nevertheless ends
in a waterfall-regime, as is typical for hybrid inflation. In such
hybrid natural inflation scenarios (examples are provided by Wilson line inflation and fluxbrane inflation), the slow-roll parameter $\epsilon$ can be sizable during an early period (relevant for the CMB spectrum). 
Subsequently, $\epsilon$ quickly becomes very small before the tachyonic instability eventually terminates the slow-roll regime. In this scenario, 
one naturally generates a considerable tensor-mode contribution in the curvature spectrum, collecting nevertheless the required amount of $e$-foldings during the final period of inflation. While non-observation of tensors by Planck is certainly not a problem, a discovery in the medium to long term future is realistic. 
\end{abstract}

\preprint{DESY-13-076, NSF-KITP-13-084}

\date{14 January 2014}

\maketitle

\section{Introduction}
Cosmological inflation~\cite{Guth:1980zm,Linde:1981mu,Albrecht:1982wi,Baumann:2009ds}, and in particular the generation of curvature perturbations through quantum fluctuations of the inflaton field~\cite{Mukhanov:1981xt,Hawking:1982cz,Starobinsky:1982ee,Guth:1982ec,Bardeen:1983qw}, allows for a surprisingly accurate description of recent Planck data~\cite{Ade:2013ktc,Ade:2013uln} (see also~\cite{Sievers:2013wk,Hinshaw:2012fq,Hou:2012xq}). Concrete field-theoretic realizations roughly fall in two classes: large-field and small-field models.

In the first class, the canonically normalized inflaton field $\phi$ covers a large distance in field space, $\Delta\phi\gg M_P$, during the last 60 $e$-folds. This has the advantage that it can be realized with a very simple potential, e.g. of the form $\sim\phi^n$~\cite{Linde:1983gd}. In addition, such models have the very interesting feature of producing sizable tensor perturbations without fine tuning. However, they require the absence or unnatural smallness of all higher-dimension operators $\sim \phi^m$ with $m>n$, which is unnatural unless an approximate shift symmetry $\phi\to\phi+c$ is present .

By contrast, small-field models work with $\Delta\phi \ll M_P$, but they require a non-generic, flat potential. Here flatness means in particular that, in spite of a significant constant term in the potential, the coefficients of the $\phi$ and $\phi^2$ terms have to be unnaturally small. In principle, this can be easily realized either by explicit tuning of the dimension-six operators~\cite{Kachru:2003sx,Baumann:2007np,Krause:2007jk,Baumann:2007ah}, or using a shift symmetry as in natural inflation\footnote{A closely related proposal is Wilson line (or `extranatural') inflation~\cite{ArkaniHamed:2003wu}.}
\cite{Freese:1990rb}, $\phi\to\phi+c$. However, for inflation to end sizable interactions of $\phi$ have to be present in some region in field space, which nevertheless leads to fine tuning.\footnote{In addition to these shift symmetry-breaking interactions the inflaton can couple shift-symmetrically to gauge fields. This coupling may provide an efficient channel for reheating~\cite{Barnaby:2011qe}.} This discussion can be taken to the more advanced level of supergravity, where it takes the form of arguing about natural sizes of higher-order terms in the K\"ahler potential~\cite{Copeland:1994vg,Linde:1997sj}. Most importantly for us, additional fine tuning is necessary to obtain large tensor perturbations.

We believe that the discussion about natural or unnatural sizes of operator coefficients can (at least in principle) be more fruitful in the context of a UV complete model, the natural candidate being of course superstring theory. Constructing large-field models in string theory has been notoriously difficult for some time. However, recently considerable progress took place based on constructions of (axion) monodromy and unwinding inflation, see e.g.~\cite{Silverstein:2008sg,McAllister:2008hb,Kaloper:2008fb,Berg:2009tg,Kaloper:2011jz,Dubovsky:2011tu,Lawrence:2012ua,D'Amico:2012sz,Shlaer:2012by,D'Amico:2012ji}. Here we focus on small-field models. As we will argue, there is an interesting class of stringy (or at least string-inspired) models, where a shift symmetry guarantees the flatness of the potential\footnote{Specific string-theoretic forms of the K\"ahler potential might also be able to do
this job, such as its simple overall volume dependence~\cite{Conlon:2005jm} or a Heisenberg symmetry~\cite{Gaillard:1995az,Antusch:2008pn,Antusch:2011ei}, but this goes beyond our present discussion.} and tensor perturbations can be sizable in spite of the small field range.

Before describing such stringy constructions, we recall the Lyth bound and our general plan for avoiding it: In slow-roll inflation the tensor-to-scalar ratio (i.e.\ the ratio of the gravitational-wave power spectrum $\Delta_{\cal T}^2\sim H^2$ and the scalar power spectrum $\Delta_{\cal R}^2\sim H^2/\epsilon$) is proportional to the first slow-roll parameter, $r = \Delta_{\cal T}^2 / \Delta_{\cal R}^2=16\epsilon$. (We use  $M_P = 1$ here and below.) This can be rewritten in terms of the field-variation per $e$-fold as
\be\label{eq:LBound}
 \sqrt{\frac{r}{8}} \simeq \left|\frac{\d \phi}{\d N}\right| \quad .
\ee
Thus, if $r$ is roughly constant or monotonically increasing during inflation, the size of tensor modes produced within the initial, observable 10 $e$-folds of the cosmologically needed 60 $e$-folds of inflation is bounded by the total field excursion during these 60 $e$-folds~\cite{Lyth:1996im,Boubekeur:2005zm}.\footnote{The bound arising from a situation of monotonically increasing $\epsilon$ during the last $60$ $e$-folds of inflation is considered in~\cite{Boubekeur:2005zm}. The original paper~\cite{Lyth:1996im} only takes into account the field excursion during the observable $\sim 10$ $e$-folds.} In particular, for small-field inflation $r$ is typically negligibly small.

In many small-field models, inflation ends because the evolution of the inflaton smoothly changes from slow-roll to fast-roll. In this sense, a monotonically increasing $\epsilon$ is a common feature. However, there are certainly exceptions: For example, if inflation ends in a `waterfall' classical instability of a second scalar field, as in hybrid inflation\footnote{Supersymmetric 
hybrid inflation incorporating various corrections was investigated in the minimal SUSY hybrid inflation program~\cite{Rehman:2009nq,Pallis:2013qz} where it was found that, in some regions of parameter space, this model can produce sizable gravity waves~\cite{Shafi:2010jr}. More generally, large tensor signals in 
small-field inflation can be obtained whenever a sufficiently complicated potential is tuned in order to achieve a non-monotonic evolution of 
$\epsilon$ (see for example~\cite{BenDayan:2009kv,Hotchkiss:2011gz}). For other ways of avoiding the Lyth bound see e.g.~\cite{Senatore:2011sp,Barnaby:2012xt,Kobayashi:2013awa,Dimastrogiovanni:2012ew}.}~\cite{Linde:1991km,Linde:1993cn}, $\epsilon$ can decrease monotonically, allowing for a large tensor signal during the observable 10 $e$-folds.

Combining the concept of a shift symmetry, protecting the flatness of the potential against radiative corrections, with the idea of hybrid inflation leads to hybrid natural inflation\footnote{The inflaton can be a pseudo-Nambu-Goldstone boson~\cite{Cohn:2000hc,Stewart:2000pa,Ross:2009hg,Ross:2010fg,Kaplan:2003aj} or a Wilson line~\cite{Kaplan:2003aj,ArkaniHamed:2003mz}. The proposed models go by various names, such as `little inflatons' or `pseudonatural inflation'. Wilson line inflation was put into a stringy context in~\cite{Avgoustidis:2006zp}. For other ideas of combining axions with hybrid inflation see e.g.~\cite{Choi:2011me,Kawasaki:2012wj}.}~\cite{Cohn:2000hc,Stewart:2000pa,Kaplan:2003aj,ArkaniHamed:2003mz,Avgoustidis:2006zp,Ross:2009hg,Ross:2010fg}. The shift symmetry is generically broken by non-perturbative effects, inducing a periodic cosine-potential for the axion. This is a rather attractive setting, as it overcomes the problematic issue~\cite{Banks:2003sx} of a super-Planckian axion decay 
constant in natural inflation (see also~\cite{Kim:2004rp,Dimopoulos:2005ac,Kallosh:2007cc,Grimm:2007hs,Berg:2009tg,Conlon:2012tz}). If, in such a model, the waterfall sets in close to the minimum of the potential, $\epsilon$ can be sizable during the observable $e$-folds of inflation, while the bulk of the required 60 $e$-folds is accumulated at later stages, when $\epsilon$ is very small. This quite naturally provides us with a potentially detectable tensor signal, in spite of the small field range. 

Motivated by the above considerations, we now focus on string theory realizations\footnote{Constructing meta-stable de Sitter vacua~\cite{Giddings:2001yu,Kachru:2003aw,Balasubramanian:2004uy,Balasubramanian:2005zx,Westphal:2006tn} and inflationary models~\cite{Dvali:1998pa,Burgess:2001fx,GarciaBellido:2001ky,Dasgupta:2002ew,Kachru:2003sx,Conlon:2005jm,Avgoustidis:2006zp,Silverstein:2008sg} in string theory is now a mature field of research.} of hybrid natural inflation. 
They arise as models of $D$-term hybrid inflation~\cite{Binetruy:1996xj,Halyo:1996pp} and can be more specifically
characterized as Wilson line inflation \cite{Avgoustidis:2006zp} on the type IIA or as fluxbrane inflation \cite{Hebecker:2011hk,Hebecker:2012aw} on the (mirror dual) type IIB side. We take the type IIB (or fluxbrane) point of view 
because moduli stabilization is better understood in this setting and because 
of the more intuitive, geometric picture of inflation: In this scenario, the inflaton is the transverse separation of a pair of D7-branes. Since the 
two branes carry opposite gauge flux, they move towards each other, which 
corresponds to the slow rolling of the inflaton. At a certain (extremely 
small) critical distance the gauge flux annihilates. This is the waterfall 
regime, with the waterfall fields realized by open strings stretched between the branes. The presence of a shift symmetry can be most easily understood 
via T-duality to type IIA: The duality maps the brane deformation modulus to a Wilson line, which is known to enjoy a shift symmetry at large volume. This shift symmetry is broken by non-perturbative effects, giving rise to the periodic (cosine shaped) potential alluded to above. 
 
For avoiding the Lyth bound, it is essential that the waterfall sets in very close to the minimum of the potential. From a purely field-theoretic perspective, this requires some tuning of Lagrangian parameters or 
an appropriate model building effort.\footnote{For example, the authors of~\cite{Ross:2009hg,Ross:2010fg} find that inflation most naturally starts and ends above the inflection point of the potential. In this regime $\epsilon$ increases monotonically and thus the Lyth bound applies, leading to a small tensor-to-scalar ratio. By contrast, the authors of \cite{Stewart:2000pa} achieve waterfall near the minimum through field theory model building, but they do not consider tensor modes.
}
From the fluxbrane perspective, however, things look different: The energy of the two-brane system is minimized in a situation where the branes come very close to each other. This is where the waterfall sets in. Moreover, while the maximal possible field excursion corresponds to roughly the length scale of the compact manifold, the critical distance of the waterfall instability is substringy~\cite{Hebecker:2011hk,WIP}. Thus, a large hierarchy between the maximal and the critical brane-to-brane distance is natural.

In this paper we discuss the phenomenology of a general hybrid inflation model with a periodic potential. In this model a sizable tensor-to-scalar ratio can be obtained for a Planckian axion decay constant. However, stringy consistency conditions dictate bounds on those decay constants. The examples of K\"ahler and complex structure axions are examined. The latter is of particular interest for us as, from an F-theory perspective, the inflaton, being a D7-brane deformation modulus, is part of the complex structure moduli space of the fourfold. We argue that for generic values of the complex structure the tensor-to-scalar ratio can be as large as $r\sim 10^{-3}$.

\section{Effective potential of fluxbrane inflation}

The effective scalar potential of fluxbrane inflation, or its T-dual cousin Wilson line inflation, can be approximated by a hybrid inflation setup with a periodic inflaton dependence from a non-perturbative correction. With $\phi,\chi$ denoting the canonically normalized inflaton and waterfall field, respectively, we have
\be\label{eq:fullPot}
V(\phi,\chi)=\frac{\lambda}{4}\,\left(\chi^2-\frac{2\Lambda^2}{\sqrt\lambda}\right)^2\cdot \left[1-\alpha\,\cos\left(\frac{\phi}{f}\right)\right]+g\phi^2\chi^2\quad.
\ee
Inflation is driven by $\phi$ in slow-roll at
\be
\phi>\phi_c\quad,\quad\phi_c^2=\frac{\sqrt\lambda}{g}\,\Lambda^2 \quad ,
\ee
with the vacuum energy dominantly sitting in the first term of \eqref{eq:fullPot} at $\chi=0$. This entails $\alpha<1$. (We choose $\alpha >0 $ by convention.) Inflation ends via a waterfall instability in $\chi$ once $\phi<\phi_c$. Keeping the vacuum energy $\Lambda^4$ during inflation fixed, we can adjust $\phi_c\ll 1$ as small as we like by choosing appropriately $\sqrt\lambda\ll g<1$.

During inflation at $\phi>\phi_c$ the dynamics is therefore governed by the effective potential
\be\label{eq:InfPot}
V(\phi)=\Lambda^4\, \left[1-\alpha\,\cos\left(\frac{\phi}{f}\right)\right]\quad.
\ee
The slow-roll parameters are~\cite{Ross:2009hg,Ross:2010fg,Kaplan:2003aj}
\bea
\epsilon&=&\frac12\left(\frac{V'}{V}\right)^2=\frac{\alpha^2}{2f^2}\,\frac{\sin^2(\phi/f)}{\big(1-\alpha\cos(\phi/f)\big)^2}\quad,\nonumber\\
\eta&=&\frac{V''}{V}=\frac{\alpha}{f^2}\,\frac{\cos(\phi/f)}{1-\alpha\cos(\phi/f)}\quad, \label{eq:SR-Param}\\
\xi^2 &=&-\,\frac{V'V'''}{V^2}=\frac{\alpha^2}{f^4}\,\frac{\sin^2(\phi/f)}{\big(1-\alpha\cos(\phi/f)\big)^2}\quad,\nonumber
\eea
from which the 2-point function observables can be computed as
\bea
n_S&=&1-6\epsilon+2\eta\quad,\nonumber\\
r&=&16\epsilon\quad,\\
\frac{{\rm d}n_S}{{\rm d}\ln k}&=& 16\epsilon\eta - 24\epsilon^2+2\xi^2 \quad.\nonumber
\eea
We see from \eqref{eq:SR-Param} that, for a given potential, maximizing $r$ means taking $\eta\to 0$. Hence, in a first step, we choose $\phi_{N}=\phi_0 \equiv \pi f/2$ to be the fixed starting point as there we have $V'$ maximal while $V''=0$.

We can now compute the number of $e$-folds $N$ elapsing between the initial field value $\phi_{N}=\phi_0$ and the final value $\phi_c$ in the limit $\phi_c/f \ll 1$:
\be\label{eq:efolds}
N=\int\limits_{t(\phi_c)}^{t(\phi_{0})}\hspace{-2ex}{\rm d}t\,H=\int\limits_{\phi_c}^{\phi_{0}}\hspace{-1ex}\frac{{\rm d}\phi}{\sqrt{2\epsilon}}\simeq \frac{f^2}{\alpha}\ln \left(\frac{4\phi_0}{\pi\phi_c}\right)  \quad .
\ee
This can be dialed by choosing the waterfall exit $\phi_c$ appropriately, so that $N = 60$, i.e.\
\be
 \phi_c \simeq \frac{4}{\pi}\phi_0  \exp\left(-\frac{60 \alpha }{f^2}\right)\quad .\label{eq:exitVal}
\ee
As $\phi_0$ is now the point at 60 $e$-folds before the end of inflation, the observables at CMB scales are evaluated at $\phi=\phi_0$. This gives the 2-point function observables
\bea\label{eq:results}
n_S&\simeq&1-6\epsilon=1-\frac38\,r\simeq 0.962+0.038\,(1-r/0.1)\quad ,\nonumber\\
\frac{{\rm d}n_S}{{\rm d}\ln k}&=&2\xi^2+{\cal O}(\alpha^4)\simeq 2\frac{\alpha^2}{f^4}  \quad ,\\
r&=&16\epsilon= 8\frac{\alpha^2}{f^2}\simeq 4 f^2 \, \frac{{\d}n_S}{{\d}\ln k}\quad.\nonumber
\eea
We now see that a choice of a Planckian axion decay constant $f=1$ and $\alpha=0.1$ produces a red tilt $n_S\simeq 0.97$ and a sizable tensor-mode fraction $r\simeq 0.08$, while keeping the running of the spectral index ${\rm d}n_S/{\rm d}\ln k\simeq 0.02$ moderately small.

However, the embedding of this effective description into a string theory model dictates additional constraints on the parameters. For our purposes, the most relevant restrictions are on axion decay constants, which are subject of section \ref{sec:Constraints}. Guided by this discussion we choose to work with the fiducial bound $f\lesssim \sqrt{3}/4\pi$. Furthermore, we have to implement the observational constraints on $n_S$ and its running. These are $n_S = 0.9603 \pm 0.0073$ and ${\rm d}n_S/{\rm d}\ln k = - 0.0134 \pm 0.0090$~\cite{Ade:2013uln}.
Using the constraints $f \lesssim \sqrt{3}/4\pi$ and $ \d n_S/ \d \ln k \lesssim 0.01$, equation~\eqref{eq:results} dictates the bound $r\lesssim 7.6 \times 10^{-4}$, which in turn forces $n_S\simeq 1$, a value excluded by Planck at the level of $5\sigma$.

However, $n_S<1$ is easily achieved by letting inflation start slightly above $\phi_0=\pi f/2$. In this region of field space $\eta$ takes the form
\be
 \eta \simeq -\frac{\alpha}{f^3}\left(\phi_{N} - \phi_0\right) \quad .
\ee
The measured value $n_S = 0.9603$ dictates $ \phi_{N}/\phi_0 \simeq 1.18 $.

Thus, we have consistently realized $r\simeq 7.6 \times 10^{-4}$ in a string-motivated setting.
By contrast, the Lyth-bound estimate of \eqref{eq:LBound}, assuming constant $r$ and $\Delta N = 60$, would give $r \simeq 1.4 \times 10^{-4}$. Hence, we gain a factor of $\simeq 5$ as compared to the Lyth approximation.

We can now compare this with the estimated precision of future cosmological probes. While the B-mode polarization search in the CMB is expected to yield sensitivity of $r=\text{(a few)}\times 10^{-2}$ for Planck~\cite{Baumann:2008aq}, the dedicated CMB polarization probe candidates CMBpol/EPIC~\cite{Baumann:2008aq} and PIXIE~\cite{Kogut:2011xw} can detect a tensor-to-scalar ratio down to $r\simeq 10^{-3}$. Even more promising is the analysis of the angular power spectra and weak lensing contribution to the 21 cm radiation, which can yield a B-mode detection down to $r\simeq 10^{-9}$~\cite{Sigurdson:2005cp,Book:2011dz}.

\subsection{Production of Primordial Black Holes}

The potential of our hybrid axion inflation model flattens towards the end of the slow-roll regime. This means that the amplitude of curvature perturbations grows, implying the threat of primordial black hole overproduction \cite{Hawking:1971ei,Carr:1974nx,Carr:1975qj}. To evaluate the situation, we compare the 
curvature perturbations $\Delta_{\cal R}^2\propto H^2/\epsilon$ at the beginning 
and the end of inflation, $\Delta_{{\cal R}, N}/ \Delta_{{\cal R}, c} \simeq \sqrt{\epsilon_c/\epsilon_{N}}$. Using~\eqref{eq:SR-Param} and~\eqref{eq:exitVal} we find
\be\label{eq:BHbound}
\frac{\Delta_{{\cal R}, N}}{\Delta_{{\cal R}, c}} \simeq 2  \exp\left(-\frac{60 \alpha }{f^2}\right) \sim 10^{-2} \quad .
\ee
Now, the most recent value \cite{Ade:2013uln} for the power spectrum 
is $\Delta_{\cal R}^2\equiv A_S \simeq 2.2 \times 10^{-9}$ at the fiducial 
scale $k = 0.05 \, \text{Mpc}^{-1}$. This can be identified with our 
$\Delta_{{\cal R}, N}^2$. It can be compared to the most conservative primordial black hole production bound, which is $\Delta_{{\cal R}, c}^2 < 10^{-3}$ (see \cite{Peiris:2008be,Josan:2009qn} and references therein). One finds $\Delta_{{\cal R}, N}/\Delta_{{\cal R}, c} \gtrsim 10^{-3} $. Thus, in 
view of \eqref{eq:BHbound}, our model is completely safe.

\subsection{Curvaton in Hybrid Natural Inflation}

Before ending this section we pause to analyze whether the curvature perturbations can be generated by an additional light scalar, i.e.\ a curvaton $\sigma$ \cite{Lyth:2001nq,Enqvist:2001zp,Moroi:2001ct} (for the specific formulae used below see \cite{Langlois:2004nn} and references therein). Due to the high friction and the small mass, the field value of the curvaton is constant during inflation $\sigma \equiv \sigma^\ast$. The time-evolution of the scalar power spectrum $\Delta_\sigma^2 \sim H^2 / (\sigma^\ast)^2$ produced by the curvaton is thus governed by the time-evolution of the Hubble parameter $H$ during inflation. Consequently, one finds that in the curvaton dominated regime ($\epsilon \gg (\sigma^\ast)^2$) the spectral index is given by
\begin{equation}
 n_S -1 = -2\epsilon .
\end{equation}
Thus, the first slow-roll parameter is bound to be $\epsilon \simeq 0.02$. Furthermore, it is clear from  \eqref{eq:SR-Param} together with the bound $f\lesssim 0.1$ that slow-roll inflation (i.e.~$\epsilon,\eta \ll 1$) can only be achieved for  $\alpha\ll 1$. Hence, in the first line of \eqref{eq:SR-Param}, we can replace $1-\alpha\cos(\phi/f)$ by 1. Together with $\epsilon\simeq 0.02$ this equation then implies $\alpha\gtrsim 0.2 f$ and therefore
\begin{equation}
 \eta \gtrsim \frac{0.2}{f}\cos\left(\frac{\phi}{f}\right) .
\end{equation}
For the fiducial value $f = \sqrt{3}/4\pi$ this gives \mbox{$\eta \gtrsim 1.4 \cdot \cos(\phi/f)$}. Now consider the running of the spectral index which, in the curvaton dominated regime, is given by
\begin{equation}\label{eq:runningCurvaton}
 \frac{{\rm d}n_S}{{\rm d}\ln k} = 4\epsilon \left(\eta - 2\epsilon\right).
\end{equation}
This value can become incompatible with the data, $|\d n_S/ \d \ln k| \lesssim 0.01$, if $\eta$ becomes large too quickly. Recall that $\eta$ is small close to the inflection point $\phi_0$. We thus have to assume $\phi_N \approx \phi_0$ in order not to get into conflict with the bound on the running from the very beginning. In analogy to \eqref{eq:efolds}, we can then derive a bound on the number $\Delta N$ of $e$-foldings which are generated while the inflaton rolls from $\phi_0$ to some smaller value $\phi$: $\Delta N \lesssim 0.7 \cdot \ln \left(\frac{4\phi_0}{\pi\phi}\right)$. (Here we have used again the fiducial value $f = \sqrt{3}/4\pi$.) It is thus clear that the inflaton leaves the region where $\cos(\phi/f)\ll 1$ already during the first $e$-fold, giving $\eta = \mathcal{O} (1)$. In view of \eqref{eq:runningCurvaton} and with $\epsilon = 0.02$, the running of the spectral index predicted by the curvaton model in hybrid natural inflation then violates the bound after one $e$-folding. Thus, taking 
the stringy bound on $f$ seriously, it is impossible to realize a curvaton-dominated power spectrum in hybrid natural inflation.

\section{Stringy Constraints}
\label{sec:Constraints}

String theory dictates additional constraints on the parameters, in particular on the axion decay constant \cite{Banks:2003sx}. Our focus will be on two types of axionic scalars: the imaginary parts of K\"ahler moduli and the real parts of complex structure moduli. K\"ahler axions descend from $p$-form potentials of the 10d theory upon dimensional reduction to 4d. To understand that complex structure moduli have anything to do with axions, recall that under mirror symmetry the complex structure moduli space of type IIB string theory is mapped to the K\"ahler moduli space of type IIA. Thus, at large complex structure (which corresponds to large volume on the type IIA side), we expect an axionic shift symmetry to act on the complex structure 
moduli as well.

As a simple example, consider the axio-dilaton $S = i/g_s + C_0$, where $g_s$ is the string coupling and $C_0$ is the Ramond-Ramond (RR) zero-form potential. The K\"ahler potential for this modulus is $K = -\ln \left(-i \left(S - \ov S\right)\right)$, giving rise to a kinetic term
\be
\mathcal L \supset K_{S \ov S} \left|\partial S \right|^2 \supset \left(\frac{g_s}{2}\right)^2 \left(\partial C_0\right)^2 \quad .
\ee
The canonically normalized `would-be' inflaton is $\phi = \frac{g_s}{\sqrt{2}}C_0$. The periodicity $C_0 \rightarrow C_0 + 1$ of the RR zero-form field~\cite{Denef:2008wq} implies a periodicity $\phi \to \phi + g_s/\sqrt{2}$, and hence $f = \frac{g_s}{\sqrt{2}\, 2\pi}$. Therefore, already at the self-dual point $g_s = 1$, the axion decay constant is much smaller than one. It decreases even further at weak coupling, $g_s \ll 1$.

From the F-theory perspective, $S$ is part of the complex structure moduli space of the fourfold. Therefore, we expect similar considerations to apply in the case of complex structure axions. The same is true for deformation moduli of D7-branes, as they are part of the complex structure of the F-theory fourfold as well. The analogs of $g_s \sim 1$ and $g_s \ll 1$ are generic and large imaginary parts of the complex structure moduli, respectively. We thus expect that the axion decay constant $f$ in \eqref{eq:InfPot} can take values as large as $f\sim 1/4\pi$ at generic complex structure and $g_s \sim 1$.

In the example of K\"ahler axions~\cite{Conlon:2006tq,Goodsell:2009xc,Cicoli:2011yh,Cicoli:2012sz} one obtains, via dimensional reduction, the term
\be\label{eq:Coupling}
 \mathcal L \supset \frac{1}{4\pi}r^{i\alpha} c_\alpha \operatorname{tr} \left(F_i \wedge F_i \right) \quad ,
\ee
which displays the coupling of the axions $c_\alpha$ to the field strength of the gauge field living on a D7-brane (wrapping a four-cycle labeled by the index $i$). The $c_\alpha$ are the coefficients of an expansion of the RR four-form in terms of a basis of four-forms (labeled by the index $\alpha$) of the threefold. The integers $r^{i \alpha}$ arise from integrating the four-form labeled by $\alpha$ over the four-cycle labeled by $i$. Quantization of $\int \operatorname{tr} \left(F_i \wedge F_i \right)$ implies that the term \eqref{eq:Coupling} is trivial for integer values of $c_\alpha$~\cite{Cicoli:2012sz}.

In order to read off the axion decay constant one has to canonically normalize the axion, using the K\"ahler potential $K = - 2\ln \mathcal V$. From here it is apparent that the axion decay constant typically scales with the inverse of some four-cycle volume, the precise value depending on the volume form. For example, for a scenario with one large four-cycle with volume\footnote{The (dimensionless) Einstein frame four-cycle volume $\tau$ is related to the string frame four-cycle volume as $\tau = g_s^{-1} \ell_s^{-4} \tau^s$, where $\ell_s = 2\pi \sqrt{\alpha'}$ is the string length.} $\tau$ and corresponding K\"ahler modulus $T = \tau + i \, c$, such that $\mathcal V \sim \left(T + \overline T\right)^{3/2}$, one finds~\cite{Cicoli:2012sz}
\be
 f = \frac{\sqrt{3}}{4\pi \tau} \quad .
\ee

We take $f\lesssim \sqrt{3}/4\pi$ as our fiducial value, corresponding to an Einstein frame four-cycle volume of unity. In the example of a compactification on a square $(T^2)^3$ and for $g_s = 1$ this is the T-self-dual point. If one instead evaluates $f$ at the point where instanton corrections $\sim e^{-2 \pi \tau}$ become important, the bound on $f$ is generally weakened by a factor of $2\pi$.

Note that for $f = \sqrt{3}/4\pi$ and $\d n_S/ \d \ln k = 0.01$ one has $\alpha = 1.3\times 10^{-3}$ which matches $\left. e^{-2\pi \tau}\right|_{\tau = 1}$ up to an $\mathcal O (1)$ factor. This is in very good agreement with our expectation that oscillatory potentials like \eqref{eq:InfPot} arise from instanton effects.

\section{Conclusions}
We have analyzed models representing a cross between axionic and hybrid inflation. They naturally produce a sufficient amount of $e$-foldings within a small field range. Moreover, the typical axionic modulation of the flat tree-level
potential leads to a variation of $\epsilon$. This generates a significant tensor-mode contribution (up to $r \sim 10^{-3}$)
in early inflation, where $\epsilon$ is sizable. The required number of $e$-foldings is accumulated later on, when $\epsilon$ approaches zero.

We have also demonstrated that, within this class of inflationary models, it is impossible to generate the curvature perturbations by a curvaton field. The crucial obstacle is the string-theoretic bound on the field range of the axion. 

Obviously, the magnitude of curvature perturbations grows
at high $\ell$. It would be interesting to look for observable consequences of this effect, e.g. along the lines of~\cite{Tashiro:2008sf,Chluba:2012gq,Dent:2012ne,Chluba:2012we}. 
Finally, a more detailed analysis of the relevant stringy construction is
clearly necessary and in progress \cite{WIP}.

\vspace*{0.5cm}
\begin{acknowledgments}
We would like to thank Stefan Sj\"ors and Timo Weigand for helpful discussions. This work was supported in part by the Impuls und Vernetzungsfond of the Helmholtz Association of German Research Centres under grant HZ-NG-603, by the Transregio TR33 
``The Dark Universe'', and by the National Science Foundation under Grant No. NSF PHY11-25915.
\end{acknowledgments}

\bibliography{Tensors}

\begin{thebibliography}{102}%
\makeatletter
\providecommand \@ifxundefined [1]{%
 \@ifx{#1\undefined}
}%
\providecommand \@ifnum [1]{%
 \ifnum #1\expandafter \@firstoftwo
 \else \expandafter \@secondoftwo
 \fi
}%
\providecommand \@ifx [1]{%
 \ifx #1\expandafter \@firstoftwo
 \else \expandafter \@secondoftwo
 \fi
}%
\providecommand \natexlab [1]{#1}%
\providecommand \enquote  [1]{``#1''}%
\providecommand \bibnamefont  [1]{#1}%
\providecommand \bibfnamefont [1]{#1}%
\providecommand \citenamefont [1]{#1}%
\providecommand \href@noop [0]{\@secondoftwo}%
\providecommand \href [0]{\begingroup \@sanitize@url \@href}%
\providecommand \@href[1]{\@@startlink{#1}\@@href}%
\providecommand \@@href[1]{\endgroup#1\@@endlink}%
\providecommand \@sanitize@url [0]{\catcode `\\12\catcode `\$12\catcode
  `\&12\catcode `\#12\catcode `\^12\catcode `\_12\catcode `\%12\relax}%
\providecommand \@@startlink[1]{}%
\providecommand \@@endlink[0]{}%
\providecommand \url  [0]{\begingroup\@sanitize@url \@url }%
\providecommand \@url [1]{\endgroup\@href {#1}{\urlprefix }}%
\providecommand \urlprefix  [0]{URL }%
\providecommand \Eprint [0]{\href }%
\providecommand \doibase [0]{http://dx.doi.org/}%
\providecommand \selectlanguage [0]{\@gobble}%
\providecommand \bibinfo  [0]{\@secondoftwo}%
\providecommand \bibfield  [0]{\@secondoftwo}%
\providecommand \translation [1]{[#1]}%
\providecommand \BibitemOpen [0]{}%
\providecommand \bibitemStop [0]{}%
\providecommand \bibitemNoStop [0]{.\EOS\space}%
\providecommand \EOS [0]{\spacefactor3000\relax}%
\providecommand \BibitemShut  [1]{\csname bibitem#1\endcsname}%
\let\auto@bib@innerbib\@empty
\bibitem [{\citenamefont {Guth}(1981)}]{Guth:1980zm}%
  \BibitemOpen
  \bibfield  {author} {\bibinfo {author} {\bibfnamefont {A.~H.}\ \bibnamefont
  {Guth}},\ }\href {\doibase 10.1103/PhysRevD.23.347} {\bibfield  {journal}
  {\bibinfo  {journal} {Phys. Rev.}\ }\textbf {\bibinfo {volume} {D23}},\
  \bibinfo {pages} {347} (\bibinfo {year} {1981})}\BibitemShut {NoStop}%
\bibitem [{\citenamefont {Linde}(1982)}]{Linde:1981mu}%
  \BibitemOpen
  \bibfield  {author} {\bibinfo {author} {\bibfnamefont {A.~D.}\ \bibnamefont
  {Linde}},\ }\href {\doibase 10.1016/0370-2693(82)91219-9} {\bibfield
  {journal} {\bibinfo  {journal} {Phys. Lett.}\ }\textbf {\bibinfo {volume}
  {B108}},\ \bibinfo {pages} {389} (\bibinfo {year} {1982})}\BibitemShut
  {NoStop}%
\bibitem [{\citenamefont {Albrecht}\ and\ \citenamefont
  {Steinhardt}(1982)}]{Albrecht:1982wi}%
  \BibitemOpen
  \bibfield  {author} {\bibinfo {author} {\bibfnamefont {A.}~\bibnamefont
  {Albrecht}}\ and\ \bibinfo {author} {\bibfnamefont {P.~J.}\ \bibnamefont
  {Steinhardt}},\ }\href {\doibase 10.1103/PhysRevLett.48.1220} {\bibfield
  {journal} {\bibinfo  {journal} {Phys. Rev. Lett.}\ }\textbf {\bibinfo
  {volume} {48}},\ \bibinfo {pages} {1220} (\bibinfo {year}
  {1982})}\BibitemShut {NoStop}%
\bibitem [{\citenamefont {Baumann}(2009)}]{Baumann:2009ds}%
  \BibitemOpen
  \bibfield  {author} {\bibinfo {author} {\bibfnamefont {D.}~\bibnamefont
  {Baumann}},\ }\href@noop {} {\  (\bibinfo {year} {2009})},\ \Eprint
  {http://arxiv.org/abs/0907.5424} {arXiv:0907.5424 [hep-th]} \BibitemShut
  {NoStop}%
\bibitem [{\citenamefont {Mukhanov}\ and\ \citenamefont
  {Chibisov}(1981)}]{Mukhanov:1981xt}%
  \BibitemOpen
  \bibfield  {author} {\bibinfo {author} {\bibfnamefont {V.~F.}\ \bibnamefont
  {Mukhanov}}\ and\ \bibinfo {author} {\bibfnamefont {G.~V.}\ \bibnamefont
  {Chibisov}},\ }\href@noop {} {\bibfield  {journal} {\bibinfo  {journal} {JETP
  Lett.}\ }\textbf {\bibinfo {volume} {33}},\ \bibinfo {pages} {532} (\bibinfo
  {year} {1981})}\BibitemShut {NoStop}%
\bibitem [{\citenamefont {Hawking}(1982)}]{Hawking:1982cz}%
  \BibitemOpen
  \bibfield  {author} {\bibinfo {author} {\bibfnamefont {S.~W.}\ \bibnamefont
  {Hawking}},\ }\href {\doibase 10.1016/0370-2693(82)90373-2} {\bibfield
  {journal} {\bibinfo  {journal} {Phys. Lett.}\ }\textbf {\bibinfo {volume}
  {B115}},\ \bibinfo {pages} {295} (\bibinfo {year} {1982})}\BibitemShut
  {NoStop}%
\bibitem [{\citenamefont {Starobinsky}(1982)}]{Starobinsky:1982ee}%
  \BibitemOpen
  \bibfield  {author} {\bibinfo {author} {\bibfnamefont {A.~A.}\ \bibnamefont
  {Starobinsky}},\ }\href {\doibase 10.1016/0370-2693(82)90541-X} {\bibfield
  {journal} {\bibinfo  {journal} {Phys. Lett.}\ }\textbf {\bibinfo {volume}
  {B117}},\ \bibinfo {pages} {175} (\bibinfo {year} {1982})}\BibitemShut
  {NoStop}%
\bibitem [{\citenamefont {Guth}\ and\ \citenamefont {Pi}(1982)}]{Guth:1982ec}%
  \BibitemOpen
  \bibfield  {author} {\bibinfo {author} {\bibfnamefont {A.~H.}\ \bibnamefont
  {Guth}}\ and\ \bibinfo {author} {\bibfnamefont {S.~Y.}\ \bibnamefont {Pi}},\
  }\href {\doibase 10.1103/PhysRevLett.49.1110} {\bibfield  {journal} {\bibinfo
   {journal} {Phys. Rev. Lett.}\ }\textbf {\bibinfo {volume} {49}},\ \bibinfo
  {pages} {1110} (\bibinfo {year} {1982})}\BibitemShut {NoStop}%
\bibitem [{\citenamefont {Bardeen}\ \emph {et~al.}(1983)\citenamefont
  {Bardeen}, \citenamefont {Steinhardt},\ and\ \citenamefont
  {Turner}}]{Bardeen:1983qw}%
  \BibitemOpen
  \bibfield  {author} {\bibinfo {author} {\bibfnamefont {J.~M.}\ \bibnamefont
  {Bardeen}}, \bibinfo {author} {\bibfnamefont {P.~J.}\ \bibnamefont
  {Steinhardt}}, \ and\ \bibinfo {author} {\bibfnamefont {M.~S.}\ \bibnamefont
  {Turner}},\ }\href {\doibase 10.1103/PhysRevD.28.679} {\bibfield  {journal}
  {\bibinfo  {journal} {Phys. Rev.}\ }\textbf {\bibinfo {volume} {D28}},\
  \bibinfo {pages} {679} (\bibinfo {year} {1983})}\BibitemShut {NoStop}%
\bibitem [{\citenamefont {Ade}\ \emph {et~al.}(2013{\natexlab{a}})\citenamefont
  {Ade} \emph {et~al.}}]{Ade:2013ktc}%
  \BibitemOpen
  \bibfield  {author} {\bibinfo {author} {\bibfnamefont {P.}~\bibnamefont
  {Ade}} \emph {et~al.} (\bibinfo {collaboration} {Planck Collaboration}),\
  }\href@noop {} {\  (\bibinfo {year} {2013}{\natexlab{a}})},\ \Eprint
  {http://arxiv.org/abs/1303.5062} {arXiv:1303.5062 [astro-ph.CO]} \BibitemShut
  {NoStop}%
\bibitem [{\citenamefont {Ade}\ \emph {et~al.}(2013{\natexlab{b}})\citenamefont
  {Ade} \emph {et~al.}}]{Ade:2013uln}%
  \BibitemOpen
  \bibfield  {author} {\bibinfo {author} {\bibfnamefont {P.}~\bibnamefont
  {Ade}} \emph {et~al.} (\bibinfo {collaboration} {Planck Collaboration}),\
  }\href@noop {} {\  (\bibinfo {year} {2013}{\natexlab{b}})},\ \Eprint
  {http://arxiv.org/abs/1303.5082} {arXiv:1303.5082 [astro-ph.CO]} \BibitemShut
  {NoStop}%
\bibitem [{\citenamefont {Sievers}\ \emph {et~al.}(2013)\citenamefont {Sievers}
  \emph {et~al.}}]{Sievers:2013wk}%
  \BibitemOpen
  \bibfield  {author} {\bibinfo {author} {\bibfnamefont {J.~L.}\ \bibnamefont
  {Sievers}} \emph {et~al.},\ }\href@noop {} {\  (\bibinfo {year} {2013})},\
  \Eprint {http://arxiv.org/abs/1301.0824} {arXiv:1301.0824 [astro-ph.CO]}
  \BibitemShut {NoStop}%
\bibitem [{\citenamefont {Hinshaw}\ \emph {et~al.}(2012)\citenamefont {Hinshaw}
  \emph {et~al.}}]{Hinshaw:2012fq}%
  \BibitemOpen
  \bibfield  {author} {\bibinfo {author} {\bibfnamefont {G.}~\bibnamefont
  {Hinshaw}} \emph {et~al.},\ }\href@noop {} {\  (\bibinfo {year} {2012})},\
  \Eprint {http://arxiv.org/abs/1212.5226} {arXiv:1212.5226 [astro-ph.CO]}
  \BibitemShut {NoStop}%
\bibitem [{\citenamefont {Hou}\ \emph {et~al.}(2012)\citenamefont {Hou} \emph
  {et~al.}}]{Hou:2012xq}%
  \BibitemOpen
  \bibfield  {author} {\bibinfo {author} {\bibfnamefont {Z.}~\bibnamefont
  {Hou}} \emph {et~al.},\ }\href@noop {} {\  (\bibinfo {year} {2012})},\
  \Eprint {http://arxiv.org/abs/1212.6267} {arXiv:1212.6267 [astro-ph.CO]}
  \BibitemShut {NoStop}%
\bibitem [{\citenamefont {Linde}(1983)}]{Linde:1983gd}%
  \BibitemOpen
  \bibfield  {author} {\bibinfo {author} {\bibfnamefont {A.~D.}\ \bibnamefont
  {Linde}},\ }\href {\doibase 10.1016/0370-2693(83)90837-7} {\bibfield
  {journal} {\bibinfo  {journal} {Phys. Lett.}\ }\textbf {\bibinfo {volume}
  {B129}},\ \bibinfo {pages} {177} (\bibinfo {year} {1983})}\BibitemShut
  {NoStop}%
\bibitem [{\citenamefont {Kachru}\ \emph
  {et~al.}(2003{\natexlab{a}})\citenamefont {Kachru} \emph
  {et~al.}}]{Kachru:2003sx}%
  \BibitemOpen
  \bibfield  {author} {\bibinfo {author} {\bibfnamefont {S.}~\bibnamefont
  {Kachru}} \emph {et~al.},\ }\href {\doibase 10.1088/1475-7516/2003/10/013}
  {\bibfield  {journal} {\bibinfo  {journal} {JCAP}\ }\textbf {\bibinfo
  {volume} {0310}},\ \bibinfo {pages} {013} (\bibinfo {year}
  {2003}{\natexlab{a}})},\ \Eprint {http://arxiv.org/abs/hep-th/0308055}
  {arXiv:hep-th/0308055} \BibitemShut {NoStop}%
\bibitem [{\citenamefont {Baumann}\ \emph {et~al.}(2007)\citenamefont
  {Baumann}, \citenamefont {Dymarsky}, \citenamefont {Klebanov}, \citenamefont
  {McAllister},\ and\ \citenamefont {Steinhardt}}]{Baumann:2007np}%
  \BibitemOpen
  \bibfield  {author} {\bibinfo {author} {\bibfnamefont {D.}~\bibnamefont
  {Baumann}}, \bibinfo {author} {\bibfnamefont {A.}~\bibnamefont {Dymarsky}},
  \bibinfo {author} {\bibfnamefont {I.~R.}\ \bibnamefont {Klebanov}}, \bibinfo
  {author} {\bibfnamefont {L.}~\bibnamefont {McAllister}}, \ and\ \bibinfo
  {author} {\bibfnamefont {P.~J.}\ \bibnamefont {Steinhardt}},\ }\href
  {\doibase 10.1103/PhysRevLett.99.141601} {\bibfield  {journal} {\bibinfo
  {journal} {Phys.Rev.Lett.}\ }\textbf {\bibinfo {volume} {99}},\ \bibinfo
  {pages} {141601} (\bibinfo {year} {2007})},\ \Eprint
  {http://arxiv.org/abs/0705.3837} {arXiv:0705.3837 [hep-th]} \BibitemShut
  {NoStop}%
\bibitem [{\citenamefont {Krause}\ and\ \citenamefont
  {Pajer}(2008)}]{Krause:2007jk}%
  \BibitemOpen
  \bibfield  {author} {\bibinfo {author} {\bibfnamefont {A.}~\bibnamefont
  {Krause}}\ and\ \bibinfo {author} {\bibfnamefont {E.}~\bibnamefont {Pajer}},\
  }\href {\doibase 10.1088/1475-7516/2008/07/023} {\bibfield  {journal}
  {\bibinfo  {journal} {JCAP}\ }\textbf {\bibinfo {volume} {0807}},\ \bibinfo
  {pages} {023} (\bibinfo {year} {2008})},\ \Eprint
  {http://arxiv.org/abs/0705.4682} {arXiv:0705.4682 [hep-th]} \BibitemShut
  {NoStop}%
\bibitem [{\citenamefont {Baumann}\ \emph {et~al.}(2008)\citenamefont
  {Baumann}, \citenamefont {Dymarsky}, \citenamefont {Klebanov},\ and\
  \citenamefont {McAllister}}]{Baumann:2007ah}%
  \BibitemOpen
  \bibfield  {author} {\bibinfo {author} {\bibfnamefont {D.}~\bibnamefont
  {Baumann}}, \bibinfo {author} {\bibfnamefont {A.}~\bibnamefont {Dymarsky}},
  \bibinfo {author} {\bibfnamefont {I.~R.}\ \bibnamefont {Klebanov}}, \ and\
  \bibinfo {author} {\bibfnamefont {L.}~\bibnamefont {McAllister}},\ }\href
  {\doibase 10.1088/1475-7516/2008/01/024} {\bibfield  {journal} {\bibinfo
  {journal} {JCAP}\ }\textbf {\bibinfo {volume} {0801}},\ \bibinfo {pages}
  {024} (\bibinfo {year} {2008})},\ \Eprint {http://arxiv.org/abs/0706.0360}
  {arXiv:0706.0360 [hep-th]} \BibitemShut {NoStop}%
\bibitem [{\citenamefont {Arkani-Hamed}\ \emph
  {et~al.}(2003{\natexlab{a}})\citenamefont {Arkani-Hamed}, \citenamefont
  {Cheng}, \citenamefont {Creminelli},\ and\ \citenamefont
  {Randall}}]{ArkaniHamed:2003wu}%
  \BibitemOpen
  \bibfield  {author} {\bibinfo {author} {\bibfnamefont {N.}~\bibnamefont
  {Arkani-Hamed}}, \bibinfo {author} {\bibfnamefont {H.-C.}\ \bibnamefont
  {Cheng}}, \bibinfo {author} {\bibfnamefont {P.}~\bibnamefont {Creminelli}}, \
  and\ \bibinfo {author} {\bibfnamefont {L.}~\bibnamefont {Randall}},\ }\href
  {\doibase 10.1103/PhysRevLett.90.221302} {\bibfield  {journal} {\bibinfo
  {journal} {Phys. Rev. Lett.}\ }\textbf {\bibinfo {volume} {90}},\ \bibinfo
  {pages} {221302} (\bibinfo {year} {2003}{\natexlab{a}})},\ \Eprint
  {http://arxiv.org/abs/hep-th/0301218} {arXiv:hep-th/0301218} \BibitemShut
  {NoStop}%
\bibitem [{\citenamefont {Freese}\ \emph {et~al.}(1990)\citenamefont {Freese},
  \citenamefont {Frieman},\ and\ \citenamefont {Olinto}}]{Freese:1990rb}%
  \BibitemOpen
  \bibfield  {author} {\bibinfo {author} {\bibfnamefont {K.}~\bibnamefont
  {Freese}}, \bibinfo {author} {\bibfnamefont {J.~A.}\ \bibnamefont {Frieman}},
  \ and\ \bibinfo {author} {\bibfnamefont {A.~V.}\ \bibnamefont {Olinto}},\
  }\href {\doibase 10.1103/PhysRevLett.65.3233} {\bibfield  {journal} {\bibinfo
   {journal} {Phys. Rev. Lett.}\ }\textbf {\bibinfo {volume} {65}},\ \bibinfo
  {pages} {3233} (\bibinfo {year} {1990})}\BibitemShut {NoStop}%
\bibitem [{\citenamefont {Barnaby}\ \emph
  {et~al.}(2012{\natexlab{a}})\citenamefont {Barnaby}, \citenamefont {Pajer},\
  and\ \citenamefont {Peloso}}]{Barnaby:2011qe}%
  \BibitemOpen
  \bibfield  {author} {\bibinfo {author} {\bibfnamefont {N.}~\bibnamefont
  {Barnaby}}, \bibinfo {author} {\bibfnamefont {E.}~\bibnamefont {Pajer}}, \
  and\ \bibinfo {author} {\bibfnamefont {M.}~\bibnamefont {Peloso}},\ }\href
  {\doibase 10.1103/PhysRevD.85.023525} {\bibfield  {journal} {\bibinfo
  {journal} {Phys.Rev.}\ }\textbf {\bibinfo {volume} {D85}},\ \bibinfo {pages}
  {023525} (\bibinfo {year} {2012}{\natexlab{a}})},\ \Eprint
  {http://arxiv.org/abs/1110.3327} {arXiv:1110.3327 [astro-ph.CO]} \BibitemShut
  {NoStop}%
\bibitem [{\citenamefont {Copeland}\ \emph {et~al.}(1994)\citenamefont
  {Copeland}, \citenamefont {Liddle}, \citenamefont {Lyth}, \citenamefont
  {Stewart},\ and\ \citenamefont {Wands}}]{Copeland:1994vg}%
  \BibitemOpen
  \bibfield  {author} {\bibinfo {author} {\bibfnamefont {E.~J.}\ \bibnamefont
  {Copeland}}, \bibinfo {author} {\bibfnamefont {A.~R.}\ \bibnamefont
  {Liddle}}, \bibinfo {author} {\bibfnamefont {D.~H.}\ \bibnamefont {Lyth}},
  \bibinfo {author} {\bibfnamefont {E.~D.}\ \bibnamefont {Stewart}}, \ and\
  \bibinfo {author} {\bibfnamefont {D.}~\bibnamefont {Wands}},\ }\href
  {\doibase 10.1103/PhysRevD.49.6410} {\bibfield  {journal} {\bibinfo
  {journal} {Phys. Rev.}\ }\textbf {\bibinfo {volume} {D49}},\ \bibinfo {pages}
  {6410} (\bibinfo {year} {1994})},\ \Eprint
  {http://arxiv.org/abs/astro-ph/9401011} {arXiv:astro-ph/9401011} \BibitemShut
  {NoStop}%
\bibitem [{\citenamefont {Linde}\ and\ \citenamefont
  {Riotto}(1997)}]{Linde:1997sj}%
  \BibitemOpen
  \bibfield  {author} {\bibinfo {author} {\bibfnamefont {A.~D.}\ \bibnamefont
  {Linde}}\ and\ \bibinfo {author} {\bibfnamefont {A.}~\bibnamefont {Riotto}},\
  }\href {\doibase 10.1103/PhysRevD.56.R1841} {\bibfield  {journal} {\bibinfo
  {journal} {Phys. Rev.}\ }\textbf {\bibinfo {volume} {D56}},\ \bibinfo {pages}
  {1841} (\bibinfo {year} {1997})},\ \Eprint
  {http://arxiv.org/abs/hep-ph/9703209} {arXiv:hep-ph/9703209} \BibitemShut
  {NoStop}%
\bibitem [{\citenamefont {Silverstein}\ and\ \citenamefont
  {Westphal}(2008)}]{Silverstein:2008sg}%
  \BibitemOpen
  \bibfield  {author} {\bibinfo {author} {\bibfnamefont {E.}~\bibnamefont
  {Silverstein}}\ and\ \bibinfo {author} {\bibfnamefont {A.}~\bibnamefont
  {Westphal}},\ }\href {\doibase 10.1103/PhysRevD.78.106003} {\bibfield
  {journal} {\bibinfo  {journal} {Phys. Rev.}\ }\textbf {\bibinfo {volume}
  {D78}},\ \bibinfo {pages} {106003} (\bibinfo {year} {2008})},\ \Eprint
  {http://arxiv.org/abs/0803.3085} {arXiv:0803.3085 [hep-th]} \BibitemShut
  {NoStop}%
\bibitem [{\citenamefont {McAllister}\ \emph {et~al.}(2010)\citenamefont
  {McAllister}, \citenamefont {Silverstein},\ and\ \citenamefont
  {Westphal}}]{McAllister:2008hb}%
  \BibitemOpen
  \bibfield  {author} {\bibinfo {author} {\bibfnamefont {L.}~\bibnamefont
  {McAllister}}, \bibinfo {author} {\bibfnamefont {E.}~\bibnamefont
  {Silverstein}}, \ and\ \bibinfo {author} {\bibfnamefont {A.}~\bibnamefont
  {Westphal}},\ }\href {\doibase 10.1103/PhysRevD.82.046003} {\bibfield
  {journal} {\bibinfo  {journal} {Phys. Rev.}\ }\textbf {\bibinfo {volume}
  {D82}},\ \bibinfo {pages} {046003} (\bibinfo {year} {2010})},\ \Eprint
  {http://arxiv.org/abs/0808.0706} {arXiv:0808.0706 [hep-th]} \BibitemShut
  {NoStop}%
\bibitem [{\citenamefont {Kaloper}\ and\ \citenamefont
  {Sorbo}(2009)}]{Kaloper:2008fb}%
  \BibitemOpen
  \bibfield  {author} {\bibinfo {author} {\bibfnamefont {N.}~\bibnamefont
  {Kaloper}}\ and\ \bibinfo {author} {\bibfnamefont {L.}~\bibnamefont
  {Sorbo}},\ }\href {\doibase 10.1103/PhysRevLett.102.121301} {\bibfield
  {journal} {\bibinfo  {journal} {Phys.Rev.Lett.}\ }\textbf {\bibinfo {volume}
  {102}},\ \bibinfo {pages} {121301} (\bibinfo {year} {2009})},\ \Eprint
  {http://arxiv.org/abs/0811.1989} {arXiv:0811.1989 [hep-th]} \BibitemShut
  {NoStop}%
\bibitem [{\citenamefont {Berg}\ \emph {et~al.}(2010)\citenamefont {Berg},
  \citenamefont {Pajer},\ and\ \citenamefont {Sjors}}]{Berg:2009tg}%
  \BibitemOpen
  \bibfield  {author} {\bibinfo {author} {\bibfnamefont {M.}~\bibnamefont
  {Berg}}, \bibinfo {author} {\bibfnamefont {E.}~\bibnamefont {Pajer}}, \ and\
  \bibinfo {author} {\bibfnamefont {S.}~\bibnamefont {Sjors}},\ }\href
  {\doibase 10.1103/PhysRevD.81.103535} {\bibfield  {journal} {\bibinfo
  {journal} {Phys.Rev.}\ }\textbf {\bibinfo {volume} {D81}},\ \bibinfo {pages}
  {103535} (\bibinfo {year} {2010})},\ \Eprint {http://arxiv.org/abs/0912.1341}
  {arXiv:0912.1341 [hep-th]} \BibitemShut {NoStop}%
\bibitem [{\citenamefont {Kaloper}\ \emph {et~al.}(2011)\citenamefont
  {Kaloper}, \citenamefont {Lawrence},\ and\ \citenamefont
  {Sorbo}}]{Kaloper:2011jz}%
  \BibitemOpen
  \bibfield  {author} {\bibinfo {author} {\bibfnamefont {N.}~\bibnamefont
  {Kaloper}}, \bibinfo {author} {\bibfnamefont {A.}~\bibnamefont {Lawrence}}, \
  and\ \bibinfo {author} {\bibfnamefont {L.}~\bibnamefont {Sorbo}},\ }\href
  {\doibase 10.1088/1475-7516/2011/03/023} {\bibfield  {journal} {\bibinfo
  {journal} {JCAP}\ }\textbf {\bibinfo {volume} {1103}},\ \bibinfo {pages}
  {023} (\bibinfo {year} {2011})},\ \Eprint {http://arxiv.org/abs/1101.0026}
  {arXiv:1101.0026 [hep-th]} \BibitemShut {NoStop}%
\bibitem [{\citenamefont {Dubovsky}\ \emph {et~al.}(2012)\citenamefont
  {Dubovsky}, \citenamefont {Lawrence},\ and\ \citenamefont
  {Roberts}}]{Dubovsky:2011tu}%
  \BibitemOpen
  \bibfield  {author} {\bibinfo {author} {\bibfnamefont {S.}~\bibnamefont
  {Dubovsky}}, \bibinfo {author} {\bibfnamefont {A.}~\bibnamefont {Lawrence}},
  \ and\ \bibinfo {author} {\bibfnamefont {M.~M.}\ \bibnamefont {Roberts}},\
  }\href {\doibase 10.1007/JHEP02(2012)053} {\bibfield  {journal} {\bibinfo
  {journal} {JHEP}\ }\textbf {\bibinfo {volume} {1202}},\ \bibinfo {pages}
  {053} (\bibinfo {year} {2012})},\ \Eprint {http://arxiv.org/abs/1105.3740}
  {arXiv:1105.3740 [hep-th]} \BibitemShut {NoStop}%
\bibitem [{\citenamefont {Lawrence}(2012)}]{Lawrence:2012ua}%
  \BibitemOpen
  \bibfield  {author} {\bibinfo {author} {\bibfnamefont {A.}~\bibnamefont
  {Lawrence}},\ }\href {\doibase 10.1103/PhysRevD.85.105029} {\bibfield
  {journal} {\bibinfo  {journal} {Phys.Rev.}\ }\textbf {\bibinfo {volume}
  {D85}},\ \bibinfo {pages} {105029} (\bibinfo {year} {2012})},\ \Eprint
  {http://arxiv.org/abs/1203.6656} {arXiv:1203.6656 [hep-th]} \BibitemShut
  {NoStop}%
\bibitem [{\citenamefont {D'Amico}\ \emph
  {et~al.}(2012{\natexlab{a}})\citenamefont {D'Amico}, \citenamefont
  {Gobbetti}, \citenamefont {Schillo},\ and\ \citenamefont
  {Kleban}}]{D'Amico:2012sz}%
  \BibitemOpen
  \bibfield  {author} {\bibinfo {author} {\bibfnamefont {G.}~\bibnamefont
  {D'Amico}}, \bibinfo {author} {\bibfnamefont {R.}~\bibnamefont {Gobbetti}},
  \bibinfo {author} {\bibfnamefont {M.}~\bibnamefont {Schillo}}, \ and\
  \bibinfo {author} {\bibfnamefont {M.}~\bibnamefont {Kleban}},\ }\href@noop {}
  {\  (\bibinfo {year} {2012}{\natexlab{a}})},\ \Eprint
  {http://arxiv.org/abs/1211.3416} {arXiv:1211.3416 [hep-th]} \BibitemShut
  {NoStop}%
\bibitem [{\citenamefont {Shlaer}(2012)}]{Shlaer:2012by}%
  \BibitemOpen
  \bibfield  {author} {\bibinfo {author} {\bibfnamefont {B.}~\bibnamefont
  {Shlaer}},\ }\href@noop {} {\  (\bibinfo {year} {2012})},\ \Eprint
  {http://arxiv.org/abs/1211.4024} {arXiv:1211.4024 [hep-th]} \BibitemShut
  {NoStop}%
\bibitem [{\citenamefont {D'Amico}\ \emph
  {et~al.}(2012{\natexlab{b}})\citenamefont {D'Amico}, \citenamefont
  {Gobbetti}, \citenamefont {Kleban},\ and\ \citenamefont
  {Schillo}}]{D'Amico:2012ji}%
  \BibitemOpen
  \bibfield  {author} {\bibinfo {author} {\bibfnamefont {G.}~\bibnamefont
  {D'Amico}}, \bibinfo {author} {\bibfnamefont {R.}~\bibnamefont {Gobbetti}},
  \bibinfo {author} {\bibfnamefont {M.}~\bibnamefont {Kleban}}, \ and\ \bibinfo
  {author} {\bibfnamefont {M.}~\bibnamefont {Schillo}},\ }\href@noop {} {\
  (\bibinfo {year} {2012}{\natexlab{b}})},\ \Eprint
  {http://arxiv.org/abs/1211.4589} {arXiv:1211.4589 [hep-th]} \BibitemShut
  {NoStop}%
\bibitem [{\citenamefont {Conlon}\ and\ \citenamefont
  {Quevedo}(2006)}]{Conlon:2005jm}%
  \BibitemOpen
  \bibfield  {author} {\bibinfo {author} {\bibfnamefont {J.~P.}\ \bibnamefont
  {Conlon}}\ and\ \bibinfo {author} {\bibfnamefont {F.}~\bibnamefont
  {Quevedo}},\ }\href {\doibase 10.1088/1126-6708/2006/01/146} {\bibfield
  {journal} {\bibinfo  {journal} {JHEP}\ }\textbf {\bibinfo {volume} {01}},\
  \bibinfo {pages} {146} (\bibinfo {year} {2006})},\ \Eprint
  {http://arxiv.org/abs/hep-th/0509012} {arXiv:hep-th/0509012} \BibitemShut
  {NoStop}%
\bibitem [{\citenamefont {Gaillard}\ \emph {et~al.}(1995)\citenamefont
  {Gaillard}, \citenamefont {Murayama},\ and\ \citenamefont
  {Olive}}]{Gaillard:1995az}%
  \BibitemOpen
  \bibfield  {author} {\bibinfo {author} {\bibfnamefont {M.~K.}\ \bibnamefont
  {Gaillard}}, \bibinfo {author} {\bibfnamefont {H.}~\bibnamefont {Murayama}},
  \ and\ \bibinfo {author} {\bibfnamefont {K.~A.}\ \bibnamefont {Olive}},\
  }\href {\doibase 10.1016/0370-2693(95)00773-E} {\bibfield  {journal}
  {\bibinfo  {journal} {Phys. Lett.}\ }\textbf {\bibinfo {volume} {B355}},\
  \bibinfo {pages} {71} (\bibinfo {year} {1995})},\ \Eprint
  {http://arxiv.org/abs/hep-ph/9504307} {arXiv:hep-ph/9504307} \BibitemShut
  {NoStop}%
\bibitem [{\citenamefont {Antusch}\ \emph {et~al.}(2009)\citenamefont
  {Antusch}, \citenamefont {Bastero-Gil}, \citenamefont {Dutta}, \citenamefont
  {King},\ and\ \citenamefont {Kostka}}]{Antusch:2008pn}%
  \BibitemOpen
  \bibfield  {author} {\bibinfo {author} {\bibfnamefont {S.}~\bibnamefont
  {Antusch}}, \bibinfo {author} {\bibfnamefont {M.}~\bibnamefont
  {Bastero-Gil}}, \bibinfo {author} {\bibfnamefont {K.}~\bibnamefont {Dutta}},
  \bibinfo {author} {\bibfnamefont {S.~F.}\ \bibnamefont {King}}, \ and\
  \bibinfo {author} {\bibfnamefont {P.~M.}\ \bibnamefont {Kostka}},\ }\href
  {\doibase 10.1088/1475-7516/2009/01/040} {\bibfield  {journal} {\bibinfo
  {journal} {JCAP}\ }\textbf {\bibinfo {volume} {0901}},\ \bibinfo {pages}
  {040} (\bibinfo {year} {2009})},\ \Eprint {http://arxiv.org/abs/0808.2425}
  {arXiv:0808.2425 [hep-ph]} \BibitemShut {NoStop}%
\bibitem [{\citenamefont {Antusch}\ \emph {et~al.}(2011)\citenamefont
  {Antusch}, \citenamefont {Dutta}, \citenamefont {Erdmenger},\ and\
  \citenamefont {Halter}}]{Antusch:2011ei}%
  \BibitemOpen
  \bibfield  {author} {\bibinfo {author} {\bibfnamefont {S.}~\bibnamefont
  {Antusch}}, \bibinfo {author} {\bibfnamefont {K.}~\bibnamefont {Dutta}},
  \bibinfo {author} {\bibfnamefont {J.}~\bibnamefont {Erdmenger}}, \ and\
  \bibinfo {author} {\bibfnamefont {S.}~\bibnamefont {Halter}},\ }\href
  {\doibase 10.1007/JHEP04(2011)065} {\bibfield  {journal} {\bibinfo  {journal}
  {JHEP}\ }\textbf {\bibinfo {volume} {04}},\ \bibinfo {pages} {065} (\bibinfo
  {year} {2011})},\ \Eprint {http://arxiv.org/abs/1102.0093} {arXiv:1102.0093
  [hep-th]} \BibitemShut {NoStop}%
\bibitem [{\citenamefont {Lyth}(1997)}]{Lyth:1996im}%
  \BibitemOpen
  \bibfield  {author} {\bibinfo {author} {\bibfnamefont {D.~H.}\ \bibnamefont
  {Lyth}},\ }\href {\doibase 10.1103/PhysRevLett.78.1861} {\bibfield  {journal}
  {\bibinfo  {journal} {Phys. Rev. Lett.}\ }\textbf {\bibinfo {volume} {78}},\
  \bibinfo {pages} {1861} (\bibinfo {year} {1997})},\ \Eprint
  {http://arxiv.org/abs/hep-ph/9606387} {arXiv:hep-ph/9606387} \BibitemShut
  {NoStop}%
\bibitem [{\citenamefont {Boubekeur}\ and\ \citenamefont
  {Lyth}(2005)}]{Boubekeur:2005zm}%
  \BibitemOpen
  \bibfield  {author} {\bibinfo {author} {\bibfnamefont {L.}~\bibnamefont
  {Boubekeur}}\ and\ \bibinfo {author} {\bibfnamefont {D.~H.}\ \bibnamefont
  {Lyth}},\ }\href {\doibase 10.1088/1475-7516/2005/07/010} {\bibfield
  {journal} {\bibinfo  {journal} {JCAP}\ }\textbf {\bibinfo {volume} {0507}},\
  \bibinfo {pages} {010} (\bibinfo {year} {2005})},\ \Eprint
  {http://arxiv.org/abs/hep-ph/0502047} {arXiv:hep-ph/0502047} \BibitemShut
  {NoStop}%
\bibitem [{\citenamefont {Rehman}\ \emph {et~al.}(2010)\citenamefont {Rehman},
  \citenamefont {Shafi},\ and\ \citenamefont {Wickman}}]{Rehman:2009nq}%
  \BibitemOpen
  \bibfield  {author} {\bibinfo {author} {\bibfnamefont {M.~U.}\ \bibnamefont
  {Rehman}}, \bibinfo {author} {\bibfnamefont {Q.}~\bibnamefont {Shafi}}, \
  and\ \bibinfo {author} {\bibfnamefont {J.~R.}\ \bibnamefont {Wickman}},\
  }\href {\doibase 10.1016/j.physletb.2009.12.010} {\bibfield  {journal}
  {\bibinfo  {journal} {Phys. Lett.}\ }\textbf {\bibinfo {volume} {B683}},\
  \bibinfo {pages} {191} (\bibinfo {year} {2010})},\ \Eprint
  {http://arxiv.org/abs/0908.3896} {arXiv:0908.3896 [hep-ph]} \BibitemShut
  {NoStop}%
\bibitem [{\citenamefont {Pallis}\ and\ \citenamefont
  {Shafi}(2013)}]{Pallis:2013qz}%
  \BibitemOpen
  \bibfield  {author} {\bibinfo {author} {\bibfnamefont {C.}~\bibnamefont
  {Pallis}}\ and\ \bibinfo {author} {\bibfnamefont {Q.}~\bibnamefont {Shafi}},\
  }\href@noop {} {\  (\bibinfo {year} {2013})},\ \Eprint
  {http://arxiv.org/abs/1304.5202} {arXiv:1304.5202 [hep-ph]} \BibitemShut
  {NoStop}%
\bibitem [{\citenamefont {Shafi}\ and\ \citenamefont
  {Wickman}(2011)}]{Shafi:2010jr}%
  \BibitemOpen
  \bibfield  {author} {\bibinfo {author} {\bibfnamefont {Q.}~\bibnamefont
  {Shafi}}\ and\ \bibinfo {author} {\bibfnamefont {J.~R.}\ \bibnamefont
  {Wickman}},\ }\href {\doibase 10.1016/j.physletb.2011.01.002} {\bibfield
  {journal} {\bibinfo  {journal} {Phys. Lett.}\ }\textbf {\bibinfo {volume}
  {B696}},\ \bibinfo {pages} {438} (\bibinfo {year} {2011})},\ \Eprint
  {http://arxiv.org/abs/1009.5340} {arXiv:1009.5340 [hep-ph]} \BibitemShut
  {NoStop}%
\bibitem [{\citenamefont {Ben-Dayan}\ and\ \citenamefont
  {Brustein}(2010)}]{BenDayan:2009kv}%
  \BibitemOpen
  \bibfield  {author} {\bibinfo {author} {\bibfnamefont {I.}~\bibnamefont
  {Ben-Dayan}}\ and\ \bibinfo {author} {\bibfnamefont {R.}~\bibnamefont
  {Brustein}},\ }\href {\doibase 10.1088/1475-7516/2010/09/007} {\bibfield
  {journal} {\bibinfo  {journal} {JCAP}\ }\textbf {\bibinfo {volume} {1009}},\
  \bibinfo {pages} {007} (\bibinfo {year} {2010})},\ \Eprint
  {http://arxiv.org/abs/0907.2384} {arXiv:0907.2384 [astro-ph.CO]} \BibitemShut
  {NoStop}%
\bibitem [{\citenamefont {Hotchkiss}\ \emph {et~al.}(2012)\citenamefont
  {Hotchkiss}, \citenamefont {Mazumdar},\ and\ \citenamefont
  {Nadathur}}]{Hotchkiss:2011gz}%
  \BibitemOpen
  \bibfield  {author} {\bibinfo {author} {\bibfnamefont {S.}~\bibnamefont
  {Hotchkiss}}, \bibinfo {author} {\bibfnamefont {A.}~\bibnamefont {Mazumdar}},
  \ and\ \bibinfo {author} {\bibfnamefont {S.}~\bibnamefont {Nadathur}},\
  }\href {\doibase 10.1088/1475-7516/2012/02/008} {\bibfield  {journal}
  {\bibinfo  {journal} {JCAP}\ }\textbf {\bibinfo {volume} {1202}},\ \bibinfo
  {pages} {008} (\bibinfo {year} {2012})},\ \Eprint
  {http://arxiv.org/abs/1110.5389} {arXiv:1110.5389 [astro-ph.CO]} \BibitemShut
  {NoStop}%
\bibitem [{\citenamefont {Senatore}\ \emph {et~al.}(2011)\citenamefont
  {Senatore}, \citenamefont {Silverstein},\ and\ \citenamefont
  {Zaldarriaga}}]{Senatore:2011sp}%
  \BibitemOpen
  \bibfield  {author} {\bibinfo {author} {\bibfnamefont {L.}~\bibnamefont
  {Senatore}}, \bibinfo {author} {\bibfnamefont {E.}~\bibnamefont
  {Silverstein}}, \ and\ \bibinfo {author} {\bibfnamefont {M.}~\bibnamefont
  {Zaldarriaga}},\ }\href@noop {} {\  (\bibinfo {year} {2011})},\ \Eprint
  {http://arxiv.org/abs/1109.0542} {arXiv:1109.0542 [hep-th]} \BibitemShut
  {NoStop}%
\bibitem [{\citenamefont {Barnaby}\ \emph
  {et~al.}(2012{\natexlab{b}})\citenamefont {Barnaby} \emph
  {et~al.}}]{Barnaby:2012xt}%
  \BibitemOpen
  \bibfield  {author} {\bibinfo {author} {\bibfnamefont {N.}~\bibnamefont
  {Barnaby}} \emph {et~al.},\ }\href {\doibase 10.1103/PhysRevD.86.103508}
  {\bibfield  {journal} {\bibinfo  {journal} {Phys. Rev.}\ }\textbf {\bibinfo
  {volume} {D86}},\ \bibinfo {pages} {103508} (\bibinfo {year}
  {2012}{\natexlab{b}})},\ \Eprint {http://arxiv.org/abs/1206.6117}
  {arXiv:1206.6117 [astro-ph.CO]} \BibitemShut {NoStop}%
\bibitem [{\citenamefont {Kobayashi}\ and\ \citenamefont
  {Takahashi}(2013)}]{Kobayashi:2013awa}%
  \BibitemOpen
  \bibfield  {author} {\bibinfo {author} {\bibfnamefont {T.}~\bibnamefont
  {Kobayashi}}\ and\ \bibinfo {author} {\bibfnamefont {T.}~\bibnamefont
  {Takahashi}},\ }\href@noop {} {\  (\bibinfo {year} {2013})},\ \Eprint
  {http://arxiv.org/abs/1303.0242} {arXiv:1303.0242 [astro-ph.CO]} \BibitemShut
  {NoStop}%
\bibitem [{\citenamefont {Dimastrogiovanni}\ and\ \citenamefont
  {Peloso}(2012)}]{Dimastrogiovanni:2012ew}%
  \BibitemOpen
  \bibfield  {author} {\bibinfo {author} {\bibfnamefont {E.}~\bibnamefont
  {Dimastrogiovanni}}\ and\ \bibinfo {author} {\bibfnamefont {M.}~\bibnamefont
  {Peloso}},\ }\href@noop {} {\  (\bibinfo {year} {2012})},\ \Eprint
  {http://arxiv.org/abs/1212.5184} {arXiv:1212.5184 [astro-ph.CO]} \BibitemShut
  {NoStop}%
\bibitem [{\citenamefont {Linde}(1991)}]{Linde:1991km}%
  \BibitemOpen
  \bibfield  {author} {\bibinfo {author} {\bibfnamefont {A.~D.}\ \bibnamefont
  {Linde}},\ }\href {\doibase 10.1016/0370-2693(91)90130-I} {\bibfield
  {journal} {\bibinfo  {journal} {Phys. Lett.}\ }\textbf {\bibinfo {volume}
  {B259}},\ \bibinfo {pages} {38} (\bibinfo {year} {1991})}\BibitemShut
  {NoStop}%
\bibitem [{\citenamefont {Linde}(1994)}]{Linde:1993cn}%
  \BibitemOpen
  \bibfield  {author} {\bibinfo {author} {\bibfnamefont {A.~D.}\ \bibnamefont
  {Linde}},\ }\href {\doibase 10.1103/PhysRevD.49.748} {\bibfield  {journal}
  {\bibinfo  {journal} {Phys. Rev.}\ }\textbf {\bibinfo {volume} {D49}},\
  \bibinfo {pages} {748} (\bibinfo {year} {1994})},\ \Eprint
  {http://arxiv.org/abs/astro-ph/9307002} {arXiv:astro-ph/9307002} \BibitemShut
  {NoStop}%
\bibitem [{\citenamefont {Cohn}\ and\ \citenamefont
  {Stewart}(2000)}]{Cohn:2000hc}%
  \BibitemOpen
  \bibfield  {author} {\bibinfo {author} {\bibfnamefont {J.~D.}\ \bibnamefont
  {Cohn}}\ and\ \bibinfo {author} {\bibfnamefont {E.~D.}\ \bibnamefont
  {Stewart}},\ }\href {\doibase 10.1016/S0370-2693(00)00089-7} {\bibfield
  {journal} {\bibinfo  {journal} {Phys. Lett.}\ }\textbf {\bibinfo {volume}
  {B475}},\ \bibinfo {pages} {231} (\bibinfo {year} {2000})},\ \Eprint
  {http://arxiv.org/abs/hep-ph/0001333} {arXiv:hep-ph/0001333} \BibitemShut
  {NoStop}%
\bibitem [{\citenamefont {Stewart}\ and\ \citenamefont
  {Cohn}(2001)}]{Stewart:2000pa}%
  \BibitemOpen
  \bibfield  {author} {\bibinfo {author} {\bibfnamefont {E.~D.}\ \bibnamefont
  {Stewart}}\ and\ \bibinfo {author} {\bibfnamefont {J.~D.}\ \bibnamefont
  {Cohn}},\ }\href {\doibase 10.1103/PhysRevD.63.083519} {\bibfield  {journal}
  {\bibinfo  {journal} {Phys. Rev.}\ }\textbf {\bibinfo {volume} {D63}},\
  \bibinfo {pages} {083519} (\bibinfo {year} {2001})},\ \Eprint
  {http://arxiv.org/abs/hep-ph/0002214} {arXiv:hep-ph/0002214} \BibitemShut
  {NoStop}%
\bibitem [{\citenamefont {Ross}\ and\ \citenamefont
  {German}(2010{\natexlab{a}})}]{Ross:2009hg}%
  \BibitemOpen
  \bibfield  {author} {\bibinfo {author} {\bibfnamefont {G.~G.}\ \bibnamefont
  {Ross}}\ and\ \bibinfo {author} {\bibfnamefont {G.}~\bibnamefont {German}},\
  }\href {\doibase 10.1016/j.physletb.2010.01.003} {\bibfield  {journal}
  {\bibinfo  {journal} {Phys. Lett.}\ }\textbf {\bibinfo {volume} {B684}},\
  \bibinfo {pages} {199} (\bibinfo {year} {2010}{\natexlab{a}})},\ \Eprint
  {http://arxiv.org/abs/0902.4676} {arXiv:0902.4676 [hep-ph]} \BibitemShut
  {NoStop}%
\bibitem [{\citenamefont {Ross}\ and\ \citenamefont
  {German}(2010{\natexlab{b}})}]{Ross:2010fg}%
  \BibitemOpen
  \bibfield  {author} {\bibinfo {author} {\bibfnamefont {G.~G.}\ \bibnamefont
  {Ross}}\ and\ \bibinfo {author} {\bibfnamefont {G.}~\bibnamefont {German}},\
  }\href {\doibase 10.1016/j.physletb.2010.06.017} {\bibfield  {journal}
  {\bibinfo  {journal} {Phys. Lett.}\ }\textbf {\bibinfo {volume} {B691}},\
  \bibinfo {pages} {117} (\bibinfo {year} {2010}{\natexlab{b}})},\ \Eprint
  {http://arxiv.org/abs/1002.0029} {arXiv:1002.0029 [hep-ph]} \BibitemShut
  {NoStop}%
\bibitem [{\citenamefont {Kaplan}\ and\ \citenamefont
  {Weiner}(2004)}]{Kaplan:2003aj}%
  \BibitemOpen
  \bibfield  {author} {\bibinfo {author} {\bibfnamefont {D.~E.}\ \bibnamefont
  {Kaplan}}\ and\ \bibinfo {author} {\bibfnamefont {N.~J.}\ \bibnamefont
  {Weiner}},\ }\href {\doibase 10.1088/1475-7516/2004/02/005} {\bibfield
  {journal} {\bibinfo  {journal} {JCAP}\ }\textbf {\bibinfo {volume} {0402}},\
  \bibinfo {pages} {005} (\bibinfo {year} {2004})},\ \Eprint
  {http://arxiv.org/abs/hep-ph/0302014} {arXiv:hep-ph/0302014} \BibitemShut
  {NoStop}%
\bibitem [{\citenamefont {Arkani-Hamed}\ \emph
  {et~al.}(2003{\natexlab{b}})\citenamefont {Arkani-Hamed}, \citenamefont
  {Cheng}, \citenamefont {Creminelli},\ and\ \citenamefont
  {Randall}}]{ArkaniHamed:2003mz}%
  \BibitemOpen
  \bibfield  {author} {\bibinfo {author} {\bibfnamefont {N.}~\bibnamefont
  {Arkani-Hamed}}, \bibinfo {author} {\bibfnamefont {H.-C.}\ \bibnamefont
  {Cheng}}, \bibinfo {author} {\bibfnamefont {P.}~\bibnamefont {Creminelli}}, \
  and\ \bibinfo {author} {\bibfnamefont {L.}~\bibnamefont {Randall}},\ }\href
  {\doibase 10.1088/1475-7516/2003/07/003} {\bibfield  {journal} {\bibinfo
  {journal} {JCAP}\ }\textbf {\bibinfo {volume} {0307}},\ \bibinfo {pages}
  {003} (\bibinfo {year} {2003}{\natexlab{b}})},\ \Eprint
  {http://arxiv.org/abs/hep-th/0302034} {arXiv:hep-th/0302034} \BibitemShut
  {NoStop}%
\bibitem [{\citenamefont {Avgoustidis}\ \emph {et~al.}(2007)\citenamefont
  {Avgoustidis}, \citenamefont {Cremades},\ and\ \citenamefont
  {Quevedo}}]{Avgoustidis:2006zp}%
  \BibitemOpen
  \bibfield  {author} {\bibinfo {author} {\bibfnamefont {A.}~\bibnamefont
  {Avgoustidis}}, \bibinfo {author} {\bibfnamefont {D.}~\bibnamefont
  {Cremades}}, \ and\ \bibinfo {author} {\bibfnamefont {F.}~\bibnamefont
  {Quevedo}},\ }\href {\doibase 10.1007/s10714-007-0454-y} {\bibfield
  {journal} {\bibinfo  {journal} {Gen. Rel. Grav.}\ }\textbf {\bibinfo {volume}
  {39}},\ \bibinfo {pages} {1203} (\bibinfo {year} {2007})},\ \Eprint
  {http://arxiv.org/abs/hep-th/0606031} {arXiv:hep-th/0606031} \BibitemShut
  {NoStop}%
\bibitem [{\citenamefont {Choi}\ and\ \citenamefont
  {Kyae}(2012)}]{Choi:2011me}%
  \BibitemOpen
  \bibfield  {author} {\bibinfo {author} {\bibfnamefont {K.-Y.}\ \bibnamefont
  {Choi}}\ and\ \bibinfo {author} {\bibfnamefont {B.}~\bibnamefont {Kyae}},\
  }\href {\doibase 10.1016/j.physletb.2011.11.045} {\bibfield  {journal}
  {\bibinfo  {journal} {Phys. Lett.}\ }\textbf {\bibinfo {volume} {B706}},\
  \bibinfo {pages} {243} (\bibinfo {year} {2012})},\ \Eprint
  {http://arxiv.org/abs/1109.4245} {arXiv:1109.4245 [astro-ph.CO]} \BibitemShut
  {NoStop}%
\bibitem [{\citenamefont {Kawasaki}\ \emph {et~al.}(2012)\citenamefont
  {Kawasaki}, \citenamefont {Kitajima},\ and\ \citenamefont
  {Nakayama}}]{Kawasaki:2012wj}%
  \BibitemOpen
  \bibfield  {author} {\bibinfo {author} {\bibfnamefont {M.}~\bibnamefont
  {Kawasaki}}, \bibinfo {author} {\bibfnamefont {N.}~\bibnamefont {Kitajima}},
  \ and\ \bibinfo {author} {\bibfnamefont {K.}~\bibnamefont {Nakayama}},\
  }\href@noop {} {\  (\bibinfo {year} {2012})},\ \Eprint
  {http://arxiv.org/abs/1211.6516} {arXiv:1211.6516 [hep-ph]} \BibitemShut
  {NoStop}%
\bibitem [{\citenamefont {Banks}\ \emph {et~al.}(2003)\citenamefont {Banks},
  \citenamefont {Dine}, \citenamefont {Fox},\ and\ \citenamefont
  {Gorbatov}}]{Banks:2003sx}%
  \BibitemOpen
  \bibfield  {author} {\bibinfo {author} {\bibfnamefont {T.}~\bibnamefont
  {Banks}}, \bibinfo {author} {\bibfnamefont {M.}~\bibnamefont {Dine}},
  \bibinfo {author} {\bibfnamefont {P.~J.}\ \bibnamefont {Fox}}, \ and\
  \bibinfo {author} {\bibfnamefont {E.}~\bibnamefont {Gorbatov}},\ }\href
  {\doibase 10.1088/1475-7516/2003/06/001} {\bibfield  {journal} {\bibinfo
  {journal} {JCAP}\ }\textbf {\bibinfo {volume} {0306}},\ \bibinfo {pages}
  {001} (\bibinfo {year} {2003})},\ \Eprint
  {http://arxiv.org/abs/hep-th/0303252} {arXiv:hep-th/0303252} \BibitemShut
  {NoStop}%
\bibitem [{\citenamefont {Kim}\ \emph {et~al.}(2005)\citenamefont {Kim},
  \citenamefont {Nilles},\ and\ \citenamefont {Peloso}}]{Kim:2004rp}%
  \BibitemOpen
  \bibfield  {author} {\bibinfo {author} {\bibfnamefont {J.~E.}\ \bibnamefont
  {Kim}}, \bibinfo {author} {\bibfnamefont {H.~P.}\ \bibnamefont {Nilles}}, \
  and\ \bibinfo {author} {\bibfnamefont {M.}~\bibnamefont {Peloso}},\ }\href
  {\doibase 10.1088/1475-7516/2005/01/005} {\bibfield  {journal} {\bibinfo
  {journal} {JCAP}\ }\textbf {\bibinfo {volume} {0501}},\ \bibinfo {pages}
  {005} (\bibinfo {year} {2005})},\ \Eprint
  {http://arxiv.org/abs/hep-ph/0409138} {arXiv:hep-ph/0409138} \BibitemShut
  {NoStop}%
\bibitem [{\citenamefont {Dimopoulos}\ \emph {et~al.}(2008)\citenamefont
  {Dimopoulos}, \citenamefont {Kachru}, \citenamefont {McGreevy},\ and\
  \citenamefont {Wacker}}]{Dimopoulos:2005ac}%
  \BibitemOpen
  \bibfield  {author} {\bibinfo {author} {\bibfnamefont {S.}~\bibnamefont
  {Dimopoulos}}, \bibinfo {author} {\bibfnamefont {S.}~\bibnamefont {Kachru}},
  \bibinfo {author} {\bibfnamefont {J.}~\bibnamefont {McGreevy}}, \ and\
  \bibinfo {author} {\bibfnamefont {J.~G.}\ \bibnamefont {Wacker}},\ }\href
  {\doibase 10.1088/1475-7516/2008/08/003} {\bibfield  {journal} {\bibinfo
  {journal} {JCAP}\ }\textbf {\bibinfo {volume} {0808}},\ \bibinfo {pages}
  {003} (\bibinfo {year} {2008})},\ \Eprint
  {http://arxiv.org/abs/hep-th/0507205} {arXiv:hep-th/0507205} \BibitemShut
  {NoStop}%
\bibitem [{\citenamefont {Kallosh}\ \emph {et~al.}(2008)\citenamefont
  {Kallosh}, \citenamefont {Sivanandam},\ and\ \citenamefont
  {Soroush}}]{Kallosh:2007cc}%
  \BibitemOpen
  \bibfield  {author} {\bibinfo {author} {\bibfnamefont {R.}~\bibnamefont
  {Kallosh}}, \bibinfo {author} {\bibfnamefont {N.}~\bibnamefont {Sivanandam}},
  \ and\ \bibinfo {author} {\bibfnamefont {M.}~\bibnamefont {Soroush}},\ }\href
  {\doibase 10.1103/PhysRevD.77.043501} {\bibfield  {journal} {\bibinfo
  {journal} {Phys. Rev.}\ }\textbf {\bibinfo {volume} {D77}},\ \bibinfo {pages}
  {043501} (\bibinfo {year} {2008})},\ \Eprint {http://arxiv.org/abs/0710.3429}
  {arXiv:0710.3429 [hep-th]} \BibitemShut {NoStop}%
\bibitem [{\citenamefont {Grimm}(2008)}]{Grimm:2007hs}%
  \BibitemOpen
  \bibfield  {author} {\bibinfo {author} {\bibfnamefont {T.~W.}\ \bibnamefont
  {Grimm}},\ }\href {\doibase 10.1103/PhysRevD.77.126007} {\bibfield  {journal}
  {\bibinfo  {journal} {Phys. Rev.}\ }\textbf {\bibinfo {volume} {D77}},\
  \bibinfo {pages} {126007} (\bibinfo {year} {2008})},\ \Eprint
  {http://arxiv.org/abs/0710.3883} {arXiv:0710.3883 [hep-th]} \BibitemShut
  {NoStop}%
\bibitem [{\citenamefont {Conlon}(2012)}]{Conlon:2012tz}%
  \BibitemOpen
  \bibfield  {author} {\bibinfo {author} {\bibfnamefont {J.~P.}\ \bibnamefont
  {Conlon}},\ }\href {\doibase 10.1088/1475-7516/2012/09/019} {\bibfield
  {journal} {\bibinfo  {journal} {JCAP}\ }\textbf {\bibinfo {volume} {1209}},\
  \bibinfo {pages} {019} (\bibinfo {year} {2012})},\ \Eprint
  {http://arxiv.org/abs/1203.5476} {arXiv:1203.5476 [hep-th]} \BibitemShut
  {NoStop}%
\bibitem [{\citenamefont {Giddings}\ \emph {et~al.}(2002)\citenamefont
  {Giddings}, \citenamefont {Kachru},\ and\ \citenamefont
  {Polchinski}}]{Giddings:2001yu}%
  \BibitemOpen
  \bibfield  {author} {\bibinfo {author} {\bibfnamefont {S.~B.}\ \bibnamefont
  {Giddings}}, \bibinfo {author} {\bibfnamefont {S.}~\bibnamefont {Kachru}}, \
  and\ \bibinfo {author} {\bibfnamefont {J.}~\bibnamefont {Polchinski}},\
  }\href {\doibase 10.1103/PhysRevD.66.106006} {\bibfield  {journal} {\bibinfo
  {journal} {Phys. Rev.}\ }\textbf {\bibinfo {volume} {D66}},\ \bibinfo {pages}
  {106006} (\bibinfo {year} {2002})},\ \Eprint
  {http://arxiv.org/abs/hep-th/0105097} {arXiv:hep-th/0105097} \BibitemShut
  {NoStop}%
\bibitem [{\citenamefont {Kachru}\ \emph
  {et~al.}(2003{\natexlab{b}})\citenamefont {Kachru}, \citenamefont {Kallosh},
  \citenamefont {Linde},\ and\ \citenamefont {Trivedi}}]{Kachru:2003aw}%
  \BibitemOpen
  \bibfield  {author} {\bibinfo {author} {\bibfnamefont {S.}~\bibnamefont
  {Kachru}}, \bibinfo {author} {\bibfnamefont {R.}~\bibnamefont {Kallosh}},
  \bibinfo {author} {\bibfnamefont {A.~D.}\ \bibnamefont {Linde}}, \ and\
  \bibinfo {author} {\bibfnamefont {S.~P.}\ \bibnamefont {Trivedi}},\ }\href
  {\doibase 10.1103/PhysRevD.68.046005} {\bibfield  {journal} {\bibinfo
  {journal} {Phys. Rev.}\ }\textbf {\bibinfo {volume} {D68}},\ \bibinfo {pages}
  {046005} (\bibinfo {year} {2003}{\natexlab{b}})},\ \Eprint
  {http://arxiv.org/abs/hep-th/0301240} {arXiv:hep-th/0301240} \BibitemShut
  {NoStop}%
\bibitem [{\citenamefont {Balasubramanian}\ and\ \citenamefont
  {Berglund}(2004)}]{Balasubramanian:2004uy}%
  \BibitemOpen
  \bibfield  {author} {\bibinfo {author} {\bibfnamefont {V.}~\bibnamefont
  {Balasubramanian}}\ and\ \bibinfo {author} {\bibfnamefont {P.}~\bibnamefont
  {Berglund}},\ }\href {\doibase 10.1088/1126-6708/2004/11/085} {\bibfield
  {journal} {\bibinfo  {journal} {JHEP}\ }\textbf {\bibinfo {volume} {11}},\
  \bibinfo {pages} {085} (\bibinfo {year} {2004})},\ \Eprint
  {http://arxiv.org/abs/hep-th/0408054} {arXiv:hep-th/0408054} \BibitemShut
  {NoStop}%
\bibitem [{\citenamefont {Balasubramanian}\ \emph {et~al.}(2005)\citenamefont
  {Balasubramanian}, \citenamefont {Berglund}, \citenamefont {Conlon},\ and\
  \citenamefont {Quevedo}}]{Balasubramanian:2005zx}%
  \BibitemOpen
  \bibfield  {author} {\bibinfo {author} {\bibfnamefont {V.}~\bibnamefont
  {Balasubramanian}}, \bibinfo {author} {\bibfnamefont {P.}~\bibnamefont
  {Berglund}}, \bibinfo {author} {\bibfnamefont {J.~P.}\ \bibnamefont
  {Conlon}}, \ and\ \bibinfo {author} {\bibfnamefont {F.}~\bibnamefont
  {Quevedo}},\ }\href {\doibase 10.1088/1126-6708/2005/03/007} {\bibfield
  {journal} {\bibinfo  {journal} {JHEP}\ }\textbf {\bibinfo {volume} {03}},\
  \bibinfo {pages} {007} (\bibinfo {year} {2005})},\ \Eprint
  {http://arxiv.org/abs/hep-th/0502058} {arXiv:hep-th/0502058} \BibitemShut
  {NoStop}%
\bibitem [{\citenamefont {Westphal}(2007)}]{Westphal:2006tn}%
  \BibitemOpen
  \bibfield  {author} {\bibinfo {author} {\bibfnamefont {A.}~\bibnamefont
  {Westphal}},\ }\href {\doibase 10.1088/1126-6708/2007/03/102} {\bibfield
  {journal} {\bibinfo  {journal} {JHEP}\ }\textbf {\bibinfo {volume} {03}},\
  \bibinfo {pages} {102} (\bibinfo {year} {2007})},\ \Eprint
  {http://arxiv.org/abs/hep-th/0611332} {arXiv:hep-th/0611332} \BibitemShut
  {NoStop}%
\bibitem [{\citenamefont {Dvali}\ and\ \citenamefont
  {Tye}(1999)}]{Dvali:1998pa}%
  \BibitemOpen
  \bibfield  {author} {\bibinfo {author} {\bibfnamefont {G.~R.}\ \bibnamefont
  {Dvali}}\ and\ \bibinfo {author} {\bibfnamefont {S.~H.~H.}\ \bibnamefont
  {Tye}},\ }\href {\doibase 10.1016/S0370-2693(99)00132-X} {\bibfield
  {journal} {\bibinfo  {journal} {Phys. Lett.}\ }\textbf {\bibinfo {volume}
  {B450}},\ \bibinfo {pages} {72} (\bibinfo {year} {1999})},\ \Eprint
  {http://arxiv.org/abs/hep-ph/9812483} {arXiv:hep-ph/9812483} \BibitemShut
  {NoStop}%
\bibitem [{\citenamefont {Burgess}\ \emph {et~al.}(2001)\citenamefont {Burgess}
  \emph {et~al.}}]{Burgess:2001fx}%
  \BibitemOpen
  \bibfield  {author} {\bibinfo {author} {\bibfnamefont {C.~P.}\ \bibnamefont
  {Burgess}} \emph {et~al.},\ }\href@noop {} {\bibfield  {journal} {\bibinfo
  {journal} {JHEP}\ }\textbf {\bibinfo {volume} {07}},\ \bibinfo {pages} {047}
  (\bibinfo {year} {2001})},\ \Eprint {http://arxiv.org/abs/hep-th/0105204}
  {arXiv:hep-th/0105204} \BibitemShut {NoStop}%
\bibitem [{\citenamefont {Garcia-Bellido}\ \emph {et~al.}(2002)\citenamefont
  {Garcia-Bellido}, \citenamefont {Rabadan},\ and\ \citenamefont
  {Zamora}}]{GarciaBellido:2001ky}%
  \BibitemOpen
  \bibfield  {author} {\bibinfo {author} {\bibfnamefont {J.}~\bibnamefont
  {Garcia-Bellido}}, \bibinfo {author} {\bibfnamefont {R.}~\bibnamefont
  {Rabadan}}, \ and\ \bibinfo {author} {\bibfnamefont {F.}~\bibnamefont
  {Zamora}},\ }\href@noop {} {\bibfield  {journal} {\bibinfo  {journal} {JHEP}\
  }\textbf {\bibinfo {volume} {01}},\ \bibinfo {pages} {036} (\bibinfo {year}
  {2002})},\ \Eprint {http://arxiv.org/abs/hep-th/0112147}
  {arXiv:hep-th/0112147} \BibitemShut {NoStop}%
\bibitem [{\citenamefont {Dasgupta}\ \emph {et~al.}(2002)\citenamefont
  {Dasgupta}, \citenamefont {Herdeiro}, \citenamefont {Hirano},\ and\
  \citenamefont {Kallosh}}]{Dasgupta:2002ew}%
  \BibitemOpen
  \bibfield  {author} {\bibinfo {author} {\bibfnamefont {K.}~\bibnamefont
  {Dasgupta}}, \bibinfo {author} {\bibfnamefont {C.}~\bibnamefont {Herdeiro}},
  \bibinfo {author} {\bibfnamefont {S.}~\bibnamefont {Hirano}}, \ and\ \bibinfo
  {author} {\bibfnamefont {R.}~\bibnamefont {Kallosh}},\ }\href {\doibase
  10.1103/PhysRevD.65.126002} {\bibfield  {journal} {\bibinfo  {journal} {Phys.
  Rev.}\ }\textbf {\bibinfo {volume} {D65}},\ \bibinfo {pages} {126002}
  (\bibinfo {year} {2002})},\ \Eprint {http://arxiv.org/abs/hep-th/0203019}
  {arXiv:hep-th/0203019} \BibitemShut {NoStop}%
\bibitem [{\citenamefont {Binetruy}\ and\ \citenamefont
  {Dvali}(1996)}]{Binetruy:1996xj}%
  \BibitemOpen
  \bibfield  {author} {\bibinfo {author} {\bibfnamefont {P.}~\bibnamefont
  {Binetruy}}\ and\ \bibinfo {author} {\bibfnamefont {G.~R.}\ \bibnamefont
  {Dvali}},\ }\href {\doibase 10.1016/S0370-2693(96)01083-0} {\bibfield
  {journal} {\bibinfo  {journal} {Phys. Lett.}\ }\textbf {\bibinfo {volume}
  {B388}},\ \bibinfo {pages} {241} (\bibinfo {year} {1996})},\ \Eprint
  {http://arxiv.org/abs/hep-ph/9606342} {arXiv:hep-ph/9606342} \BibitemShut
  {NoStop}%
\bibitem [{\citenamefont {Halyo}(1996)}]{Halyo:1996pp}%
  \BibitemOpen
  \bibfield  {author} {\bibinfo {author} {\bibfnamefont {E.}~\bibnamefont
  {Halyo}},\ }\href {\doibase 10.1016/0370-2693(96)01001-5} {\bibfield
  {journal} {\bibinfo  {journal} {Phys. Lett.}\ }\textbf {\bibinfo {volume}
  {B387}},\ \bibinfo {pages} {43} (\bibinfo {year} {1996})},\ \Eprint
  {http://arxiv.org/abs/hep-ph/9606423} {arXiv:hep-ph/9606423} \BibitemShut
  {NoStop}%
\bibitem [{\citenamefont {Hebecker}\ \emph {et~al.}(2012)\citenamefont
  {Hebecker}, \citenamefont {Kraus}, \citenamefont {L{\"u}st}, \citenamefont
  {Steinfurt},\ and\ \citenamefont {Weigand}}]{Hebecker:2011hk}%
  \BibitemOpen
  \bibfield  {author} {\bibinfo {author} {\bibfnamefont {A.}~\bibnamefont
  {Hebecker}}, \bibinfo {author} {\bibfnamefont {S.~C.}\ \bibnamefont {Kraus}},
  \bibinfo {author} {\bibfnamefont {D.}~\bibnamefont {L{\"u}st}}, \bibinfo
  {author} {\bibfnamefont {S.}~\bibnamefont {Steinfurt}}, \ and\ \bibinfo
  {author} {\bibfnamefont {T.}~\bibnamefont {Weigand}},\ }\href {\doibase
  10.1016/j.nuclphysb.2011.08.025} {\bibfield  {journal} {\bibinfo  {journal}
  {Nucl. Phys.}\ }\textbf {\bibinfo {volume} {B854}},\ \bibinfo {pages} {509}
  (\bibinfo {year} {2012})},\ \Eprint {http://arxiv.org/abs/1104.5016}
  {arXiv:1104.5016 [hep-th]} \BibitemShut {NoStop}%
\bibitem [{\citenamefont {Hebecker}\ \emph {et~al.}(2013)\citenamefont
  {Hebecker}, \citenamefont {Kraus}, \citenamefont {K{\"u}ntzler},
  \citenamefont {L{\"u}st},\ and\ \citenamefont {Weigand}}]{Hebecker:2012aw}%
  \BibitemOpen
  \bibfield  {author} {\bibinfo {author} {\bibfnamefont {A.}~\bibnamefont
  {Hebecker}}, \bibinfo {author} {\bibfnamefont {S.~C.}\ \bibnamefont {Kraus}},
  \bibinfo {author} {\bibfnamefont {M.}~\bibnamefont {K{\"u}ntzler}}, \bibinfo
  {author} {\bibfnamefont {D.}~\bibnamefont {L{\"u}st}}, \ and\ \bibinfo
  {author} {\bibfnamefont {T.}~\bibnamefont {Weigand}},\ }\href {\doibase
  10.1007/JHEP01(2013)095} {\bibfield  {journal} {\bibinfo  {journal} {JHEP}\
  }\textbf {\bibinfo {volume} {01}},\ \bibinfo {pages} {095} (\bibinfo {year}
  {2013})},\ \Eprint {http://arxiv.org/abs/1207.2766} {arXiv:1207.2766
  [hep-th]} \BibitemShut {NoStop}%
\bibitem [{\citenamefont {Arends}\ \emph {et~al.}(2013)\citenamefont {Arends},
  \citenamefont {Hebecker}, \citenamefont {Heimpel}, \citenamefont {Kraus},
  \citenamefont {L{\"u}st}, \citenamefont {Mayrhofer}, \citenamefont {Schick},\
  and\ \citenamefont {Weigand}}]{WIP}%
  \BibitemOpen
  \bibfield  {author} {\bibinfo {author} {\bibfnamefont {M.}~\bibnamefont
  {Arends}}, \bibinfo {author} {\bibfnamefont {A.}~\bibnamefont {Hebecker}},
  \bibinfo {author} {\bibfnamefont {K.}~\bibnamefont {Heimpel}}, \bibinfo
  {author} {\bibfnamefont {S.}~\bibnamefont {Kraus}}, \bibinfo {author}
  {\bibfnamefont {D.}~\bibnamefont {L{\"u}st}}, \bibinfo {author}
  {\bibfnamefont {C.}~\bibnamefont {Mayrhofer}}, \bibinfo {author}
  {\bibfnamefont {C.}~\bibnamefont {Schick}}, \ and\ \bibinfo {author}
  {\bibfnamefont {T.}~\bibnamefont {Weigand}},\ }\href@noop {} {\bibfield
  {journal} {\bibinfo  {journal} {(work in progress)}\ } (\bibinfo {year}
  {2013})}\BibitemShut {NoStop}%
\bibitem [{\citenamefont {Baumann}\ \emph {et~al.}(2009)\citenamefont {Baumann}
  \emph {et~al.}}]{Baumann:2008aq}%
  \BibitemOpen
  \bibfield  {author} {\bibinfo {author} {\bibfnamefont {D.}~\bibnamefont
  {Baumann}} \emph {et~al.} (\bibinfo {collaboration} {CMBPol Study Team}),\
  }\href {\doibase 10.1063/1.3160885} {\bibfield  {journal} {\bibinfo
  {journal} {AIP Conf.Proc.}\ }\textbf {\bibinfo {volume} {1141}},\ \bibinfo
  {pages} {10} (\bibinfo {year} {2009})},\ \Eprint
  {http://arxiv.org/abs/0811.3919} {arXiv:0811.3919 [astro-ph]} \BibitemShut
  {NoStop}%
\bibitem [{\citenamefont {Kogut}\ \emph {et~al.}(2011)\citenamefont {Kogut},
  \citenamefont {Fixsen}, \citenamefont {Chuss}, \citenamefont {Dotson},
  \citenamefont {Dwek} \emph {et~al.}}]{Kogut:2011xw}%
  \BibitemOpen
  \bibfield  {author} {\bibinfo {author} {\bibfnamefont {A.}~\bibnamefont
  {Kogut}}, \bibinfo {author} {\bibfnamefont {D.}~\bibnamefont {Fixsen}},
  \bibinfo {author} {\bibfnamefont {D.}~\bibnamefont {Chuss}}, \bibinfo
  {author} {\bibfnamefont {J.}~\bibnamefont {Dotson}}, \bibinfo {author}
  {\bibfnamefont {E.}~\bibnamefont {Dwek}},  \emph {et~al.},\ }\href {\doibase
  10.1088/1475-7516/2011/07/025} {\bibfield  {journal} {\bibinfo  {journal}
  {JCAP}\ }\textbf {\bibinfo {volume} {1107}},\ \bibinfo {pages} {025}
  (\bibinfo {year} {2011})},\ \Eprint {http://arxiv.org/abs/1105.2044}
  {arXiv:1105.2044 [astro-ph.CO]} \BibitemShut {NoStop}%
\bibitem [{\citenamefont {Sigurdson}\ and\ \citenamefont
  {Cooray}(2005)}]{Sigurdson:2005cp}%
  \BibitemOpen
  \bibfield  {author} {\bibinfo {author} {\bibfnamefont {K.}~\bibnamefont
  {Sigurdson}}\ and\ \bibinfo {author} {\bibfnamefont {A.}~\bibnamefont
  {Cooray}},\ }\href {\doibase 10.1103/PhysRevLett.95.211303} {\bibfield
  {journal} {\bibinfo  {journal} {Phys. Rev. Lett.}\ }\textbf {\bibinfo
  {volume} {95}},\ \bibinfo {pages} {211303} (\bibinfo {year} {2005})},\
  \Eprint {http://arxiv.org/abs/astro-ph/0502549} {arXiv:astro-ph/0502549}
  \BibitemShut {NoStop}%
\bibitem [{\citenamefont {Book}\ \emph {et~al.}(2012)\citenamefont {Book},
  \citenamefont {Kamionkowski},\ and\ \citenamefont {Schmidt}}]{Book:2011dz}%
  \BibitemOpen
  \bibfield  {author} {\bibinfo {author} {\bibfnamefont {L.}~\bibnamefont
  {Book}}, \bibinfo {author} {\bibfnamefont {M.}~\bibnamefont {Kamionkowski}},
  \ and\ \bibinfo {author} {\bibfnamefont {F.}~\bibnamefont {Schmidt}},\ }\href
  {\doibase 10.1103/PhysRevLett.108.211301} {\bibfield  {journal} {\bibinfo
  {journal} {Phys.Rev.Lett.}\ }\textbf {\bibinfo {volume} {108}},\ \bibinfo
  {pages} {211301} (\bibinfo {year} {2012})},\ \Eprint
  {http://arxiv.org/abs/1112.0567} {arXiv:1112.0567 [astro-ph.CO]} \BibitemShut
  {NoStop}%
\bibitem [{\citenamefont {Hawking}(1971)}]{Hawking:1971ei}%
  \BibitemOpen
  \bibfield  {author} {\bibinfo {author} {\bibfnamefont {S.}~\bibnamefont
  {Hawking}},\ }\href@noop {} {\bibfield  {journal} {\bibinfo  {journal} {Mon.
  Not. Roy. Astron. Soc.}\ }\textbf {\bibinfo {volume} {152}},\ \bibinfo
  {pages} {75} (\bibinfo {year} {1971})}\BibitemShut {NoStop}%
\bibitem [{\citenamefont {Carr}\ and\ \citenamefont
  {Hawking}(1974)}]{Carr:1974nx}%
  \BibitemOpen
  \bibfield  {author} {\bibinfo {author} {\bibfnamefont {B.~J.}\ \bibnamefont
  {Carr}}\ and\ \bibinfo {author} {\bibfnamefont {S.~W.}\ \bibnamefont
  {Hawking}},\ }\href@noop {} {\bibfield  {journal} {\bibinfo  {journal} {Mon.
  Not. Roy. Astron. Soc.}\ }\textbf {\bibinfo {volume} {168}},\ \bibinfo
  {pages} {399} (\bibinfo {year} {1974})}\BibitemShut {NoStop}%
\bibitem [{\citenamefont {Carr}(1975)}]{Carr:1975qj}%
  \BibitemOpen
  \bibfield  {author} {\bibinfo {author} {\bibfnamefont {B.~J.}\ \bibnamefont
  {Carr}},\ }\href {\doibase 10.1086/153853} {\bibfield  {journal} {\bibinfo
  {journal} {Astrophys. J.}\ }\textbf {\bibinfo {volume} {201}},\ \bibinfo
  {pages} {1} (\bibinfo {year} {1975})}\BibitemShut {NoStop}%
\bibitem [{\citenamefont {Peiris}\ and\ \citenamefont
  {Easther}(2008)}]{Peiris:2008be}%
  \BibitemOpen
  \bibfield  {author} {\bibinfo {author} {\bibfnamefont {H.~V.}\ \bibnamefont
  {Peiris}}\ and\ \bibinfo {author} {\bibfnamefont {R.}~\bibnamefont
  {Easther}},\ }\href {\doibase 10.1088/1475-7516/2008/07/024} {\bibfield
  {journal} {\bibinfo  {journal} {JCAP}\ }\textbf {\bibinfo {volume} {0807}},\
  \bibinfo {pages} {024} (\bibinfo {year} {2008})},\ \Eprint
  {http://arxiv.org/abs/0805.2154} {arXiv:0805.2154 [astro-ph]} \BibitemShut
  {NoStop}%
\bibitem [{\citenamefont {Josan}\ \emph {et~al.}(2009)\citenamefont {Josan},
  \citenamefont {Green},\ and\ \citenamefont {Malik}}]{Josan:2009qn}%
  \BibitemOpen
  \bibfield  {author} {\bibinfo {author} {\bibfnamefont {A.~S.}\ \bibnamefont
  {Josan}}, \bibinfo {author} {\bibfnamefont {A.~M.}\ \bibnamefont {Green}}, \
  and\ \bibinfo {author} {\bibfnamefont {K.~A.}\ \bibnamefont {Malik}},\ }\href
  {\doibase 10.1103/PhysRevD.79.103520} {\bibfield  {journal} {\bibinfo
  {journal} {Phys. Rev.}\ }\textbf {\bibinfo {volume} {D79}},\ \bibinfo {pages}
  {103520} (\bibinfo {year} {2009})},\ \Eprint {http://arxiv.org/abs/0903.3184}
  {arXiv:0903.3184 [astro-ph.CO]} \BibitemShut {NoStop}%
\bibitem [{\citenamefont {Lyth}\ and\ \citenamefont
  {Wands}(2002)}]{Lyth:2001nq}%
  \BibitemOpen
  \bibfield  {author} {\bibinfo {author} {\bibfnamefont {D.~H.}\ \bibnamefont
  {Lyth}}\ and\ \bibinfo {author} {\bibfnamefont {D.}~\bibnamefont {Wands}},\
  }\href {\doibase 10.1016/S0370-2693(01)01366-1} {\bibfield  {journal}
  {\bibinfo  {journal} {Phys.Lett.}\ }\textbf {\bibinfo {volume} {B524}},\
  \bibinfo {pages} {5} (\bibinfo {year} {2002})},\ \Eprint
  {http://arxiv.org/abs/hep-ph/0110002} {arXiv:hep-ph/0110002 [hep-ph]}
  \BibitemShut {NoStop}%
\bibitem [{\citenamefont {Enqvist}\ and\ \citenamefont
  {Sloth}(2002)}]{Enqvist:2001zp}%
  \BibitemOpen
  \bibfield  {author} {\bibinfo {author} {\bibfnamefont {K.}~\bibnamefont
  {Enqvist}}\ and\ \bibinfo {author} {\bibfnamefont {M.~S.}\ \bibnamefont
  {Sloth}},\ }\href {\doibase 10.1016/S0550-3213(02)00043-3} {\bibfield
  {journal} {\bibinfo  {journal} {Nucl.Phys.}\ }\textbf {\bibinfo {volume}
  {B626}},\ \bibinfo {pages} {395} (\bibinfo {year} {2002})},\ \Eprint
  {http://arxiv.org/abs/hep-ph/0109214} {arXiv:hep-ph/0109214 [hep-ph]}
  \BibitemShut {NoStop}%
\bibitem [{\citenamefont {Moroi}\ and\ \citenamefont
  {Takahashi}(2001)}]{Moroi:2001ct}%
  \BibitemOpen
  \bibfield  {author} {\bibinfo {author} {\bibfnamefont {T.}~\bibnamefont
  {Moroi}}\ and\ \bibinfo {author} {\bibfnamefont {T.}~\bibnamefont
  {Takahashi}},\ }\href {\doibase 10.1016/S0370-2693(01)01295-3} {\bibfield
  {journal} {\bibinfo  {journal} {Phys.Lett.}\ }\textbf {\bibinfo {volume}
  {B522}},\ \bibinfo {pages} {215} (\bibinfo {year} {2001})},\ \Eprint
  {http://arxiv.org/abs/hep-ph/0110096} {arXiv:hep-ph/0110096 [hep-ph]}
  \BibitemShut {NoStop}%
\bibitem [{\citenamefont {Langlois}\ and\ \citenamefont
  {Vernizzi}(2004)}]{Langlois:2004nn}%
  \BibitemOpen
  \bibfield  {author} {\bibinfo {author} {\bibfnamefont {D.}~\bibnamefont
  {Langlois}}\ and\ \bibinfo {author} {\bibfnamefont {F.}~\bibnamefont
  {Vernizzi}},\ }\href {\doibase 10.1103/PhysRevD.70.063522} {\bibfield
  {journal} {\bibinfo  {journal} {Phys.Rev.}\ }\textbf {\bibinfo {volume}
  {D70}},\ \bibinfo {pages} {063522} (\bibinfo {year} {2004})},\ \Eprint
  {http://arxiv.org/abs/astro-ph/0403258} {arXiv:astro-ph/0403258 [astro-ph]}
  \BibitemShut {NoStop}%
\bibitem [{\citenamefont {Denef}(2008)}]{Denef:2008wq}%
  \BibitemOpen
  \bibfield  {author} {\bibinfo {author} {\bibfnamefont {F.}~\bibnamefont
  {Denef}},\ }\href@noop {} {\  (\bibinfo {year} {2008})},\ \Eprint
  {http://arxiv.org/abs/0803.1194} {arXiv:0803.1194 [hep-th]} \BibitemShut
  {NoStop}%
\bibitem [{\citenamefont {Conlon}(2006)}]{Conlon:2006tq}%
  \BibitemOpen
  \bibfield  {author} {\bibinfo {author} {\bibfnamefont {J.~P.}\ \bibnamefont
  {Conlon}},\ }\href {\doibase 10.1088/1126-6708/2006/05/078} {\bibfield
  {journal} {\bibinfo  {journal} {JHEP}\ }\textbf {\bibinfo {volume} {05}},\
  \bibinfo {pages} {078} (\bibinfo {year} {2006})},\ \Eprint
  {http://arxiv.org/abs/hep-th/0602233} {arXiv:hep-th/0602233} \BibitemShut
  {NoStop}%
\bibitem [{\citenamefont {Goodsell}\ \emph {et~al.}(2009)\citenamefont
  {Goodsell}, \citenamefont {J{\"a}ckel}, \citenamefont {Redondo},\ and\
  \citenamefont {Ringwald}}]{Goodsell:2009xc}%
  \BibitemOpen
  \bibfield  {author} {\bibinfo {author} {\bibfnamefont {M.}~\bibnamefont
  {Goodsell}}, \bibinfo {author} {\bibfnamefont {J.}~\bibnamefont
  {J{\"a}ckel}}, \bibinfo {author} {\bibfnamefont {J.}~\bibnamefont {Redondo}},
  \ and\ \bibinfo {author} {\bibfnamefont {A.}~\bibnamefont {Ringwald}},\
  }\href {\doibase 10.1088/1126-6708/2009/11/027} {\bibfield  {journal}
  {\bibinfo  {journal} {JHEP}\ }\textbf {\bibinfo {volume} {11}},\ \bibinfo
  {pages} {027} (\bibinfo {year} {2009})},\ \Eprint
  {http://arxiv.org/abs/0909.0515} {arXiv:0909.0515 [hep-ph]} \BibitemShut
  {NoStop}%
\bibitem [{\citenamefont {Cicoli}\ \emph {et~al.}(2011)\citenamefont {Cicoli},
  \citenamefont {Goodsell}, \citenamefont {J{\"a}ckel},\ and\ \citenamefont
  {Ringwald}}]{Cicoli:2011yh}%
  \BibitemOpen
  \bibfield  {author} {\bibinfo {author} {\bibfnamefont {M.}~\bibnamefont
  {Cicoli}}, \bibinfo {author} {\bibfnamefont {M.}~\bibnamefont {Goodsell}},
  \bibinfo {author} {\bibfnamefont {J.}~\bibnamefont {J{\"a}ckel}}, \ and\
  \bibinfo {author} {\bibfnamefont {A.}~\bibnamefont {Ringwald}},\ }\href
  {\doibase 10.1007/JHEP07(2011)114} {\bibfield  {journal} {\bibinfo  {journal}
  {JHEP}\ }\textbf {\bibinfo {volume} {07}},\ \bibinfo {pages} {114} (\bibinfo
  {year} {2011})},\ \Eprint {http://arxiv.org/abs/1103.3705} {arXiv:1103.3705
  [hep-th]} \BibitemShut {NoStop}%
\bibitem [{\citenamefont {Cicoli}\ \emph {et~al.}(2012)\citenamefont {Cicoli},
  \citenamefont {Goodsell}, \citenamefont {Ringwald}, \citenamefont
  {Goodsell},\ and\ \citenamefont {Ringwald}}]{Cicoli:2012sz}%
  \BibitemOpen
  \bibfield  {author} {\bibinfo {author} {\bibfnamefont {M.}~\bibnamefont
  {Cicoli}}, \bibinfo {author} {\bibfnamefont {M.}~\bibnamefont {Goodsell}},
  \bibinfo {author} {\bibfnamefont {A.}~\bibnamefont {Ringwald}}, \bibinfo
  {author} {\bibfnamefont {M.}~\bibnamefont {Goodsell}}, \ and\ \bibinfo
  {author} {\bibfnamefont {A.}~\bibnamefont {Ringwald}},\ }\href {\doibase
  10.1007/JHEP10(2012)146} {\bibfield  {journal} {\bibinfo  {journal} {JHEP}\
  }\textbf {\bibinfo {volume} {10}},\ \bibinfo {pages} {146} (\bibinfo {year}
  {2012})},\ \Eprint {http://arxiv.org/abs/1206.0819} {arXiv:1206.0819
  [hep-th]} \BibitemShut {NoStop}%
\bibitem [{\citenamefont {Tashiro}\ and\ \citenamefont
  {Sugiyama}(2008)}]{Tashiro:2008sf}%
  \BibitemOpen
  \bibfield  {author} {\bibinfo {author} {\bibfnamefont {H.}~\bibnamefont
  {Tashiro}}\ and\ \bibinfo {author} {\bibfnamefont {N.}~\bibnamefont
  {Sugiyama}},\ }\href {\doibase 10.1103/PhysRevD.78.023004} {\bibfield
  {journal} {\bibinfo  {journal} {Phys. Rev.}\ }\textbf {\bibinfo {volume}
  {D78}},\ \bibinfo {pages} {023004} (\bibinfo {year} {2008})},\ \Eprint
  {http://arxiv.org/abs/0801.3172} {arXiv:0801.3172 [astro-ph]} \BibitemShut
  {NoStop}%
\bibitem [{\citenamefont {Chluba}\ \emph
  {et~al.}(2012{\natexlab{a}})\citenamefont {Chluba}, \citenamefont {Khatri},\
  and\ \citenamefont {Sunyaev}}]{Chluba:2012gq}%
  \BibitemOpen
  \bibfield  {author} {\bibinfo {author} {\bibfnamefont {J.}~\bibnamefont
  {Chluba}}, \bibinfo {author} {\bibfnamefont {R.}~\bibnamefont {Khatri}}, \
  and\ \bibinfo {author} {\bibfnamefont {R.~A.}\ \bibnamefont {Sunyaev}},\
  }\href@noop {} {\  (\bibinfo {year} {2012}{\natexlab{a}})},\ \Eprint
  {http://arxiv.org/abs/1202.0057} {arXiv:1202.0057 [astro-ph.CO]} \BibitemShut
  {NoStop}%
\bibitem [{\citenamefont {Dent}\ \emph {et~al.}(2012)\citenamefont {Dent},
  \citenamefont {Easson},\ and\ \citenamefont {Tashiro}}]{Dent:2012ne}%
  \BibitemOpen
  \bibfield  {author} {\bibinfo {author} {\bibfnamefont {J.~B.}\ \bibnamefont
  {Dent}}, \bibinfo {author} {\bibfnamefont {D.~A.}\ \bibnamefont {Easson}}, \
  and\ \bibinfo {author} {\bibfnamefont {H.}~\bibnamefont {Tashiro}},\ }\href
  {\doibase 10.1103/PhysRevD.86.023514} {\bibfield  {journal} {\bibinfo
  {journal} {Phys. Rev.}\ }\textbf {\bibinfo {volume} {D86}},\ \bibinfo {pages}
  {023514} (\bibinfo {year} {2012})},\ \Eprint {http://arxiv.org/abs/1202.6066}
  {arXiv:1202.6066 [astro-ph.CO]} \BibitemShut {NoStop}%
\bibitem [{\citenamefont {Chluba}\ \emph
  {et~al.}(2012{\natexlab{b}})\citenamefont {Chluba}, \citenamefont
  {Erickcek},\ and\ \citenamefont {Ben-Dayan}}]{Chluba:2012we}%
  \BibitemOpen
  \bibfield  {author} {\bibinfo {author} {\bibfnamefont {J.}~\bibnamefont
  {Chluba}}, \bibinfo {author} {\bibfnamefont {A.~L.}\ \bibnamefont
  {Erickcek}}, \ and\ \bibinfo {author} {\bibfnamefont {I.}~\bibnamefont
  {Ben-Dayan}},\ }\href {\doibase 10.1088/0004-637X/758/2/76} {\bibfield
  {journal} {\bibinfo  {journal} {Astrophys.J.}\ }\textbf {\bibinfo {volume}
  {758}},\ \bibinfo {pages} {76} (\bibinfo {year} {2012}{\natexlab{b}})},\
  \Eprint {http://arxiv.org/abs/1203.2681} {arXiv:1203.2681 [astro-ph.CO]}
  \BibitemShut {NoStop}%
\end{thebibliography}%

\end{document}